\documentclass[aps,prl,reprint,showpacs,superscriptaddress,longbibliography]{revtex4-2}
\usepackage[english]{babel}
\usepackage{amsmath,amssymb,bbm,graphicx,color,comment,txfonts}
\usepackage[bookmarks=true,colorlinks,citecolor=blue,urlcolor=blue]{hyperref}
\usepackage{dsfont}
\usepackage{soul}

\usepackage{bbm}
\usepackage{float}

\usepackage{dcolumn}   
\usepackage{bm}        
\usepackage{filecontents}
\usepackage{lineno}

\usepackage{mathtools}
\usepackage[usenames,dvipsnames]{xcolor} 

\usepackage[export]{adjustbox}
\usepackage{adjustbox}

\usepackage{braket}
\usepackage{tikz}
\usepackage{bbold}

\newcommand{\bs}[1]{\boldsymbol{#1}}

\let\baraccent=\= \renewcommand{\=}[1]{\stackrel{#1}{=}}
\newcommand{\eq}[1]{Eq.\thinspace(\ref{#1})}
\newcommand{\fig}[1]{Fig.\thinspace{}\ref{#1}}
\newcommand{\fc}[1]{({#1})}
\newcommand{\figc}[2]{Fig.\thinspace{}\ref{#1}\thinspace{}\fc{#2}}

\usepackage[normalem]{ulem}

\begin{document}
\title{Critically slow operator dynamics in constrained many-body systems}
\author{Johannes Feldmeier}
\affiliation{Department of Physics and Institute for Advanced Study, Technical University of Munich, 85748 Garching, Germany}
\affiliation{Munich Center for Quantum Science and Technology (MCQST), Schellingstr. 4, D-80799 M{\"u}nchen, Germany}

\author{Michael Knap}
\affiliation{Department of Physics and Institute for Advanced Study, Technical University of Munich, 85748 Garching, Germany}
\affiliation{Munich Center for Quantum Science and Technology (MCQST), Schellingstr. 4, D-80799 M{\"u}nchen, Germany}
\date{\today}

\begin{abstract}
The far-from-equilibrium dynamics of generic interacting quantum systems is characterized by a handful of universal guiding principles, among them the ballistic spreading of initially local operators. Here, we show that in certain constrained many-body systems the structure of conservation laws can cause a drastic modification of this universal behavior. As an example, we study operator growth characterized by out-of-time-order correlations (OTOCs) in a dipole-conserving fracton chain. We identify a critical point with sub-ballistically moving OTOC front, that separates a ballistic from a dynamically frozen phase. This critical point is tied to an underlying localization transition and we use its associated scaling properties to derive an effective description of the moving operator front via a biased random walk with long waiting times. We support our arguments numerically using classically simulable automaton circuits.
\end{abstract}

\maketitle

\textbf{\textit{Introduction.}}--
The characterization of interacting quantum many-body systems out of equilibrium is at the forefront of current theoretical and experimental efforts~\cite{Deutsch91,Srednicki94,Rigol2008,alessio2016_chaos,kaufman2016_thermalization,Brydges2019_renyi}.
In recent years, immense progress has been made by focusing on \textit{universal} nonequilibrium processes in thermalizing chaotic systems, including transport~\cite{chaikin_lubensky_1995,Mukerjee06,Lux14,Bohrdt16}, entanglement growth~\cite{nahum2017_entanglement,jonay2018_coarse,knap2018_scrambling,
rakovszky2019_renyi,rakovszky2019_entanglement}, and operator spreading~\cite{nahum2018_operator,Keyserlingk2018,khemani2018_operator,Rakovszky18,nahum2018_griffith,Parker19}. The key characteristics of these processes are captured by effective hydrodynamic descriptions at long wavelengths, rendering them qualitatively independent of microscopic details~\cite{nahum2017_entanglement,nahum2018_operator,Keyserlingk2018}. As such, they can be captured in minimal models of random unitary circuits, which often prove to be simpler to handle -- both analytically and numerically -- than microscopic Hamiltonians.
The central ingredient entering such models is the structure of their conservation laws, which qualitatively affects the aforementioned processes. Usual charge conservation, for example, gives rise to emergent diffusive transport~\cite{chaikin_lubensky_1995,Mukerjee06,Lux14,Bohrdt16}. Moreover, the ballistically moving front of out-of-time-order correlators (OTOCs), which characterize the spatial spreading of operators, is augmented by algebraic tails caused by charge conservation~\cite{khemani2018_operator,Rakovszky18}.

Recently, a class of constrained fracton systems~\cite{nandkishore2019_fractons,pretko2020_fracton,
chamon2005_glass,haah2011_code,yoshida2013_fractal,vijay2015_topo,Vijay16,pretko2018_elasticity} that conserve both a charge and its associated dipole moment~\cite{pretko2017_subdim,pretko2018_gaugprinciple,pretko_witten,williamson2019_fractonic} have been shown -- theoretically and in quantum simulation experiments -- to exhibit unconventional transport properties, such as localization~\cite{Sala19,khemani20192d,Rakovszky20,scherg2021_kinetic} and subdiffusive relaxation~\cite{gromov2020_fractonhydro,feldmeier2020anomalous,
morningstar2020_kinetic,zhang_2020,moudgalya2021_spectral,Guardado20}. The former is due to a `strong fragmentation' of the Hilbert space into disjoint subsectors~\cite{Sala19,khemani20192d} while the latter is a more generic consequence of dipole conservation for transport~\cite{gromov2020_fractonhydro,feldmeier2020anomalous}. Given the impact of charge conservation on OTOCs, these observations raise the interesting question whether also the universal aspects of operator dynamics may be modified in the presence of constraints.

\begin{figure}[t]
\centering
\includegraphics[trim={0cm 0cm 0cm 0cm},clip,width=0.99\linewidth]{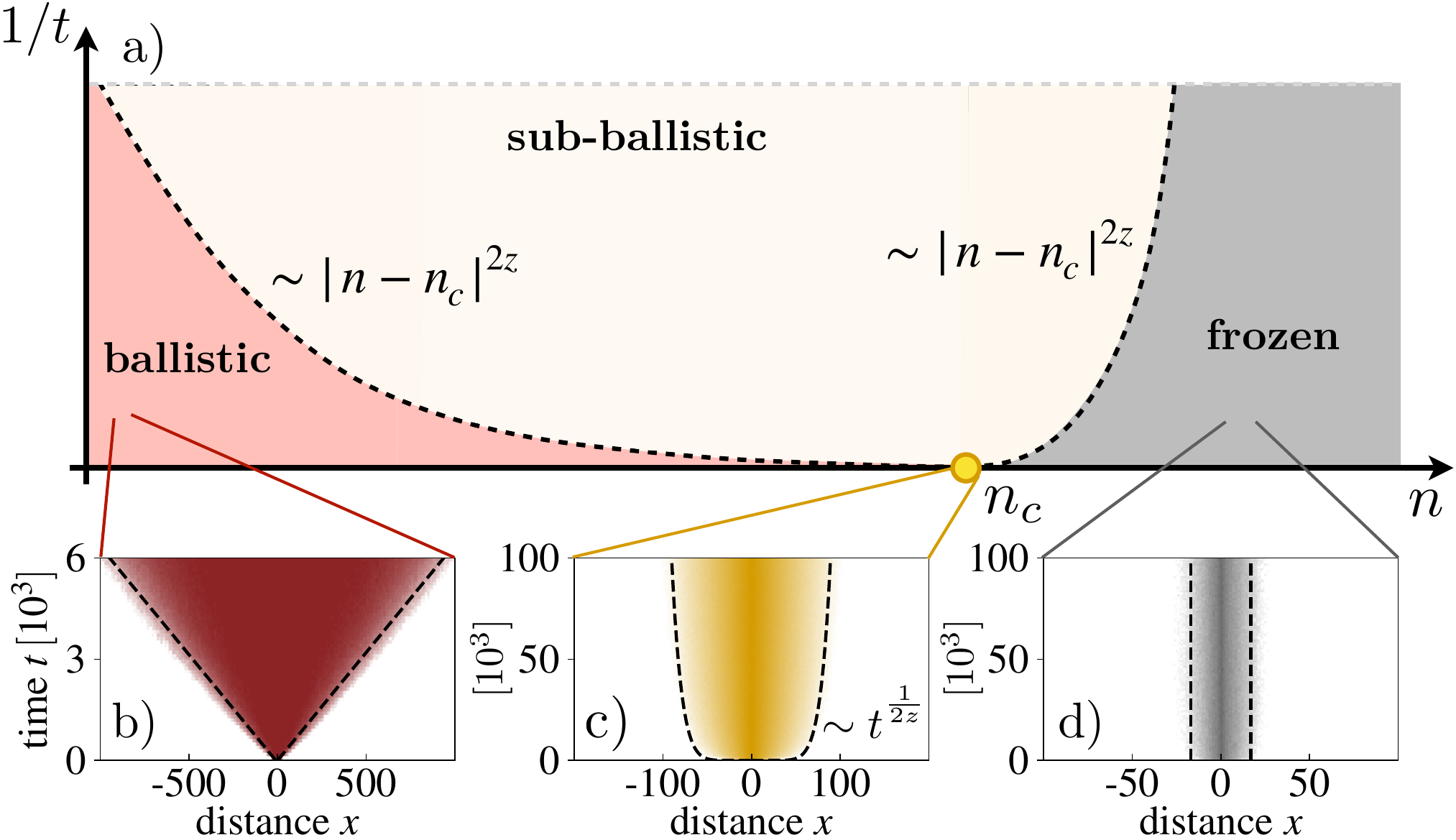}
\caption{\textbf{Dynamical phase diagram of operator spreading.} 
Our dipole conserving system exhibits a localization transition at a critical density $n_c$.
In the limit of long time scales, operators spread ballistically on the ergodic side $n < n_c$, and are frozen on the non-ergodic side $n>n_c$ of the transition. At the critical density, the operator front spreads sub-ballistically as $\sim t^{1/2z}$ with $z=4$.  The lower panels show the associated OTOC; \eq{eq:3}. The sub-ballistic spread is visible also away from the critical density at finite times within a `critical fan', bounded by a crossover time that diverges as $|n-n_c|^{-2z}$ upon approaching the transition.}
\label{fig:1}
\end{figure}

In this work, we find such new behavior in the operator dynamics of a one-dimensional, dipole-conserving chain. As demonstrated in Ref.~\cite{morningstar2020_kinetic}, such systems can be tuned from an ergodic phase with subdiffusive transport to a localized phase as a function of density. We use classically simulable automaton circuits~\cite{gopalakrishnan2018_operator,Gopalakrishnan_2018,Iaconis19,iaconis2021_complexity,iaconis2020_measurement} to determine the spread of the operator front  in these different regimes. While the OTOC front propagates ballistically at long times in the ergodic phase and freezes in the localized phase, we find the critical point between the two phases to be characterized by a \textit{sub-ballistic} spread, see \fig{fig:1}. We provide a phenomenological model that explains these  numerical results via the effects of long-lived and localized rare regions, leading to a description of the moving front in terms of a biased random walk with long waiting times.\\

\textbf{\textit{Model and circuit evolution.}}--
We focus on a one-dimensional chain of length $L$ with $3$-state onsite Hilbert space. The computational basis states are denoted by $\ket{\bs{n}} = \ket{(n_{-L/2},...,n_{L/2-1})}$, where $n_x \in \{0,1,2\}$ labels the local qutrit basis. Defining the occupation number operators $\hat{Z}_x$ via $\hat{Z}_x\ket{\bs{n}}=n_x\ket{\bs{n}}$, we can construct a global $U(1)$ charge $\hat{Q} = \sum_{x} \hat{Z}_x$ as well as its associated dipole moment $\hat{P} = \sum_{x} x\, \hat{Z}_x$. In the following, we are interested in unitary time evolution operators $\hat{U}(t)$ that commute with both $\hat{Q}$ and $\hat{P}$ simultaneously. We decompose $\hat{U}(t)  = \prod_{i} \hat{U}_i$ in terms of a layered circuit structure of local $4$-site gates $\hat{U}_i$, see \fig{fig:4}. The local gates $\hat{U}_i$ in turn are chosen randomly, subjected to the condition of preserving $\hat{Q}$ and $\hat{P}$. As selecting the $\hat{U}_i$'s from the full set of Haar random unitary gates proves challenging for large-scale numerical simulations, we work with randomly chosen \textit{automaton} gates~\cite{gopalakrishnan2018_operator,Gopalakrishnan_2018,Iaconis19,iaconis2021_complexity}, which satisfy the condition $\hat{U}_i \ket{\bs{n}} = e^{i\theta_{\bs{n}}} \ket{\bs{n}^\prime}$.
Thus, product states within the computational basis $\ket{\bs{n}}$ are, up to a phase $\theta_{\bs{n}}$, mapped to new product states within that basis. Such circuits can effectively be simulated as classical cellular automata and are valuable tools to capture universal dynamical properties at infinite temperature.

Concretely, such random circuits were recently used to demonstrate subdiffusive transport in the above model~\cite{feldmeier2020anomalous,morningstar2020_kinetic,
iaconis2021_multipole,feng2021_casimir}, as observed in the (connected) charge correlations:
\begin{equation} \label{eq:2}
\overline{\braket{\hat{Z}_x(t)\, \hat{Z}_0(0)}}_{n=1} -1 \sim t^{-1/z}F(x^z/t).
\end{equation}
Here, $z=4$ is the dynamical transport exponent, $F(\cdot)$ is a universal scaling function, $\overline{\cdot\cdot\cdot}$ is a circuit average and $\braket{\cdot}_{n}$ denotes an average over initial states at a chemical potential $\mu(n)$ that fixes an average charge density $n$. 
Furthermore, Ref.~\cite{morningstar2020_kinetic} showed the existence of a \textit{localization} transition at a critical density close to (and possibly exactly at) $n_c = 1.5$. Due to particle-hole symmetry there is another critical density $\tilde{n}_c = 0.5$. Here, we restrict our analysis to $1\leq n \leq 2$ and work with $n_c=1.5$ exactly in the analysis of our numerical results.
For $n>n_c$ the charge correlations $\overline{\braket{\hat{Z}_x(t)\, \hat{Z}_0(0)}}_{n>n_c}  - n^2 \rightarrow \mathrm{const.} \neq 0$ no longer decay towards their equilibrium value.\\

\textbf{\textit{Out-of-time-order correlators.}}--
We now proceed to study the spreading of operators. This can be done via out-of-time-ordered correlation functions (OTOCs). In particular, we consider the correlations
\begin{equation} \label{eq:3}
C^{(n)}_{ZX}(x,t) = \overline{\Braket{[\hat{Z}_x(t),\hat{X}_0(0)][\hat{Z}_x(t),\hat{X}_0(0)]^\dagger}}_{n},
\end{equation}
where we have defined a shift operator $\hat{X}_0 = \ket{2}_0\bra{0} + \ket{0}_0\bra{1} + \ket{1}_0\bra{2}$ that modifies the charge at site $0$. Using the automaton evolution introduced in the previous section, the OTOC of \eq{eq:3} takes the convenient form
\begin{align} \label{eq:4}
C^{(n)}_{ZX}(x,t) &= \sum_{\bs{n}} \frac{e^{-\mu \sum_x n_x}}{Z_{n}} \, \overline{\Bigl[ \braket{\bs{n}|\hat{Z}_x(t)|\bs{n}} - \bra{\bs{n}}\hat{X}_0 \, \hat{Z}_x(t)\, \hat{X}_0^\dagger\ket{\bs{n}} \Bigr]^2} \nonumber \\
&= \sum_{\bs{n}} \frac{e^{-\mu \sum_x n_x}}{Z_{n}} \, \overline{C^{}_{ZX}(x,t;\bs{n})} \, ,
\end{align}
where we introduced a 'single-shot' OTOC $C^{}_{ZX}(x,t;\bs{n})$ for a single initial state $\ket{\bs{n}}$ and a single circuit realization. This quantity is efficiently evaluated numerically and amounts to the local charge difference at site $x$ between two initial states $\ket{\bs{n}}$, $\hat{X}_0^\dagger\ket{\bs{n}}$ that are evolved in parallel up to time $t$ via the same circuit realization. The remaining sum over initial states $\ket{\bs{n}}$ in \eq{eq:4} can be sampled stochastically. We emphasize that the OTOC $C^{(n)}_{ZX}(x,t)$ for automaton time evolutions has been shown to reproduce the expected features of more general random unitary circuit structures~\cite{gopalakrishnan2018_operator,Iaconis19,chen2020_slow,liu2021_butterfly}.

\begin{figure}[t]
\centering
\includegraphics[trim={0cm 0cm 0cm 0cm},clip,width=0.87\linewidth]{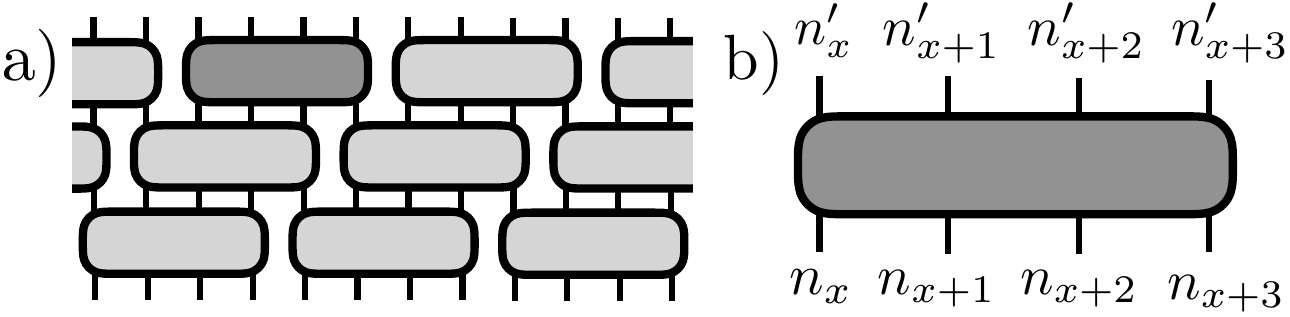}
\caption{\textbf{Time evolution.} \textbf{a)} The time evolution is given by a layered circuit structure consisting of random unitary automaton gates of range four. \textbf{b)} The gates map local strings of occupation numbers to other such strings under the constraints of charge- and dipole-conservation.}
\label{fig:4}
\end{figure}

Let us first focus on the qualitative dynamics of the OTOC, \eq{eq:3}, at long times for different limiting cases: Deep in the ergodic phase at half-filling $n=1$, \figc{fig:1}{b} shows the common light-cone structure, with a ballistically propagating OTOC front. A detailed numerical analysis of the full OTOC shape also confirms a diffusive broadening $\sim \sqrt{t}$ of the front, as well as the presence of algebraic tails $\sim (v_B t - x)^{-1/4}$ behind the front due to slow hydrodynamic transport (see supplemental material~\cite{suppmat}, Sec. A).
Deep in the localized regime $n>n_c$, we instead find the OTOC freezes in time with an exponential envelope in space, see \figc{fig:1}{d} and~\cite{suppmat}, Sec. A. This agrees with a recent numerical study in a strongly fragmented Heisenberg ladder~\cite{hahn2021_information}. The length scale on which the OTOC freezes is given by the correlation length $\xi\sim |n-n_c|^{-2}$~\cite{morningstar2020_kinetic} upon approaching the critical point from above. 
Finally, at the critical density $n=n_c=1.5$, \figc{fig:1}{c} indicates a \textit{sub-ballistic} spreading of the OTOC front. This last point is an intriguing result: What is the mechanism behind the sub-ballistic spread? Can we understand the sub-ballistic exponent that describes the moving front? And how does this critical property affect the dynamics away from $n=n_c$?

To answer these questions, we consider the single-shot OTOCs $C^{}_{ZX}(x,t;\bs{n})$ of \eq{eq:4}, see \figc{fig:2}{a} for a typical example at $n=1$. Importantly, if we denote by $x_r(t)$ the rightmost site where $C^{}_{ZX}(x,t;\bs{n}) \neq 0$ in \figc{fig:2}{a}, we notice the presence of long waiting times $\tau$ that slow the propagation of the OTOC. Thus, in order to understand the long-time dynamics of the OTOC front in terms of the propagating boundary $\overline{\braket{x_r(t)}}$, we require to understand the origin and the probability distribution $p_n(\tau)$ of these waiting times.\\

\textbf{\textit{Phenomenological model.}}--
Let us consider a toy example that illustrates the presence of waiting times in the spreading of the OTOC:  Take a small patch of seven fully filled sites, described by a state $\ket{m} = \ket{2\, 2\, 2\, 2\, 2\, 2\, 2}$. Upon inserting e.g. the operator $\hat{X}_0$ at the central site, we obtain the state $\hat{X}_0\ket{m} = \ket{2\, 2\, 2\, 1\, 2\, 2\, 2}$. There are now \textit{no} dipole-conserving, 4-site gates acting within the patch that can change this charge distribution. This observation crucially depends on dipole-conservation: For charge-conservation only, one could always randomly shift the site with reduced charge, leading, for generic systems, to a diffusively spreading OTOC within 
this patch~\cite{Rakovszky18}. Here, instead, the OTOC is `trapped' inside the localized region, implying a waiting time $\tau$ that lasts until the localized  region is breached from the \textit{outside}.

We can develop this argument more systematically as depicted in \figc{fig:2}{b}: We assume a localized region of length $\ell$ in which the average charge density exceeds the critical value, i.e. $\sum_{x=x_0}^{x_0+\ell-1} (n_x - n_c) > 0$, embedded in a system with average density below the critical value. We then expect that an operator inserted into this region implies an OTOC that is effectively `stuck' until the localized region has melted from the outside. The latter is a transport process that is governed by the subdiffusive exponent $z=4$ of \eq{eq:2}, and we therefore predict the mean waiting time associated to a localized region of length $\ell$ to be $\tau = \tau (\ell) \sim \ell^z = \ell^4$. For the purpose of this argument, we assume that the prefactor in this relation stays finite for all background densities at and below $n_c$. Now, given $\tau(\ell)$, we are required to determine the density-dependent distribution $p_n(\ell)$ which yields the probability of a localized region having length $\ell$. This can be achieved using an argument based on random walk theory: Consider a small localized region of length $\ell^\prime$ with $N(\ell^\prime) := \sum_{x=x_0}^{x_0+\ell^\prime-1} (n_x - n_c) > 0$. If we increase this region by one site to the right, we obtain $N(\ell^\prime +1)=N(\ell^\prime)+ n-n_c + \delta$, where $n$ is the average density of the system and $\delta$ is a random variable with mean zero. Therefore, $N(\ell^\prime)$ performs a biased random walk and we obtain a closed, localized region of length $\ell^\prime$ when $N(\ell^\prime)=0$ crosses zero. We are thus interested in the probability distribution of the lengths $\ell$ of first passage of zero, which assumes the asymptotic form~\cite{feller2008_probability} $p_n(\ell) \sim \ell^{-3/2}\exp(-\ell/\ell_c)$. Herein, $\ell_c \sim |n-n_c|^{-2}$ close to the transition, i.e. the system `sees' the non-critical value of the density only above a length scale where typical density fluctuations $\delta n \sim 1/\sqrt{\ell}$ of a region of length $\ell$ are comparable to $|n-n_c|$.
Equipped with the distribution $p_n(\ell)$ of localized regions and the waiting time $\tau(\ell)$ associated to such regions, we can make a qualitative estimate for the probability distribution $p_n(\tau)$ of waiting times by setting $p_n(\tau) \, d\tau = p_n(\ell(\tau)) \, d\ell$, which yields
\begin{equation} \label{eq:5}
\begin{split}
p_n(\tau) \sim \tau^{- (1 + 1/2z)} \exp\bigl\{-(\tau / \tau_c)^{1/z}\bigr\},
\end{split}
\end{equation}
where
\begin{equation} \label{eq:6}
\tau_c \sim |n-n_c|^{-2z}, \qquad z = 4.
\end{equation}
\eq{eq:5} and \eq{eq:6}, valid for $n \leq n_c$, are the central predictions of this work. From the tail of this waiting time distribution~\cite{shlesinger1974_random}, which enters the biased walk performed by the OTOC boundary $x_r(t)$, and from the localization of the system for $n>n_c$~\cite{morningstar2020_kinetic} , we can predict the long-time dynamics of the OTOC front
\begin{equation} \label{eq:7}
\overline{\braket{x_r(t)}} \sim t^\alpha \qquad\text{with} \qquad
\begin{cases}
\alpha = 1, \; n < n_c \\
\alpha = \frac{1}{2z}, \; n = n_c \\
\alpha = 0, \; n > n_c
\end{cases}.
\end{equation}
In particular, \eq{eq:7} agrees with the limiting cases studied in \figc{fig:1}{b-d}, where a sub-ballistic spread with exponent $1/2z$ separates a ballistic and a frozen long-time regime.\\

\begin{figure}[t]
\centering
\includegraphics[trim={0cm 0cm 0cm 0cm},clip,width=0.92\linewidth]{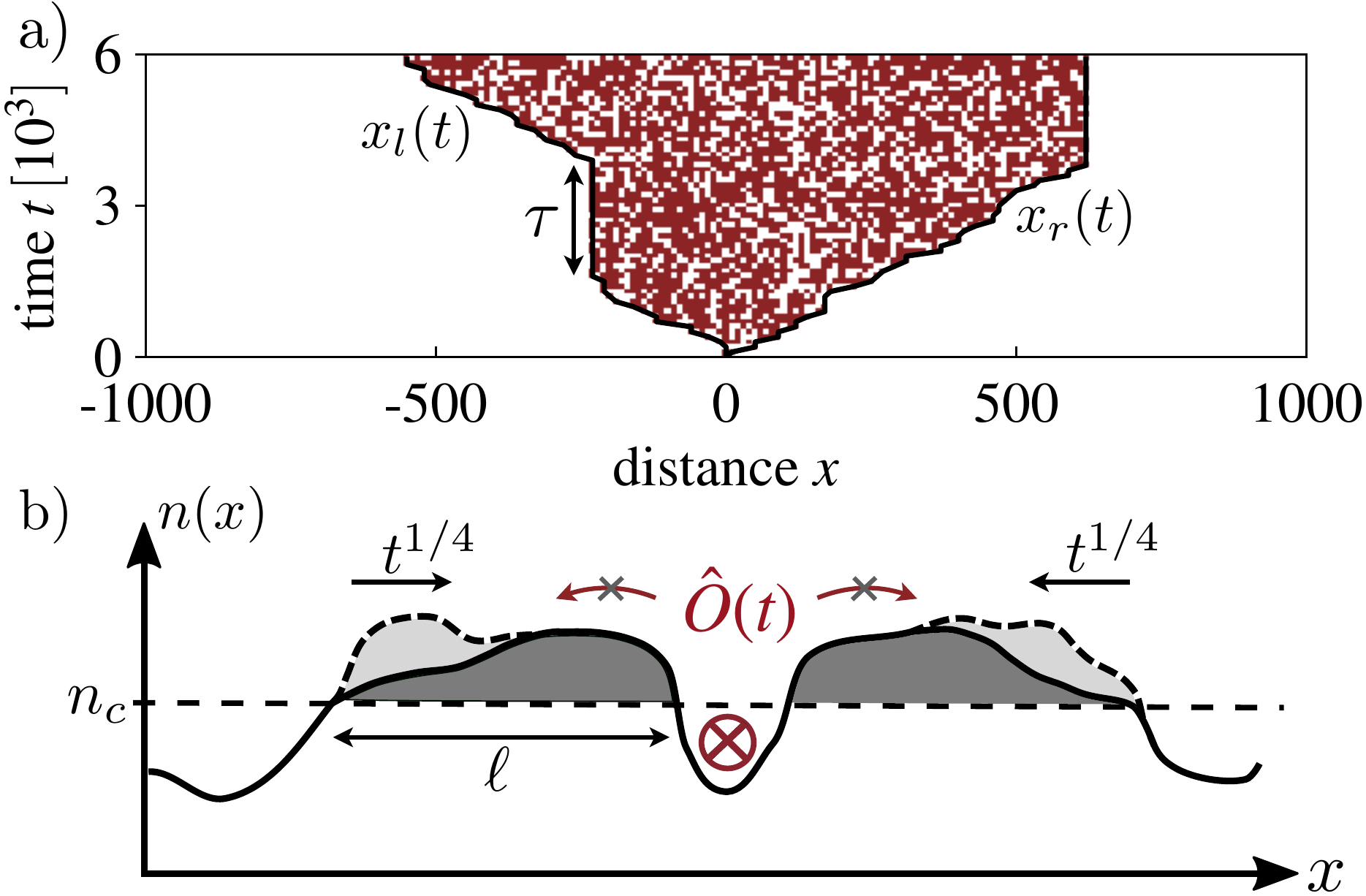}
\caption{\textbf{Origin of waiting times for operator spreading.} \textbf{a)} The single-shot OTOC, shown for a randomly chosen initial state at half-filling $n=1.0$, experiences waiting times that slow its spread.
\textbf{b)} A generic state chosen at an average density $n < n_c$ locally features regions of size $\ell$ whose density exceeds the critical value $n_c$. An operator inserted into such a localized region cannot escape from within -- a property inherent to the fractonic constraint of dipole-conservation. Instead, the operator is subject to a waiting time $\tau \sim \ell^z=\ell^4$ set by the time required for transport to melt the frozen region from the outside.
}
\label{fig:2}
\end{figure}

\begin{figure*}[t]
\centering
\includegraphics[trim={0cm 0cm 0cm 0cm},clip,width=0.99\linewidth]{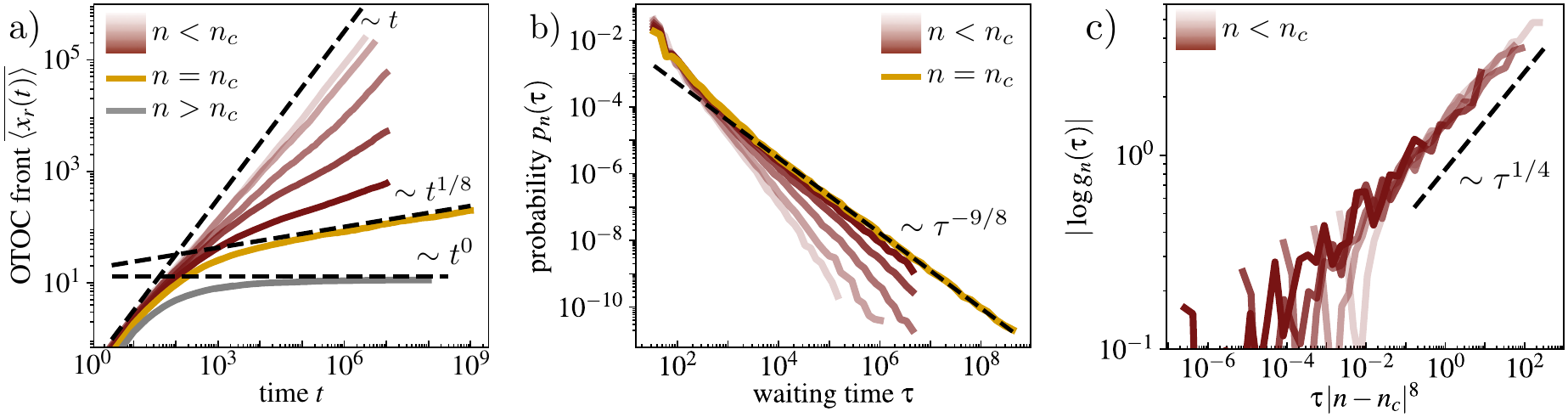}
\caption{\textbf{OTOC front and waiting time distributions.} \textbf{a)} Average $\overline{\braket{x_r(t)}}$ of the position of the OTOC front as defined in \figc{fig:2}{b}. For $n<n_c$, an upward bending towards an expected ballistic growth can be observed on the simulated timescales for densities deep in the ergodic phase. At the critical density $n_c=1.5$, the dynamics of the front shows the sub-ballistic power law $\overline{\braket{x_r(t)}}\sim t^{1/8}$, and the spread remains bounded $\braket{x_r(t)}<\mathrm{const.}$ in the localized phase. The densities shown are $n=1.0 - 1.5$ in steps of $0.1$ (light to dark red), as well as $n=1.7$. \textbf{b)} Probability distributions $p_n(\tau)$ of waiting times in the automaton circuit dynamics evaluated at densities below and at $n_c$ (same values as in \textbf{a)}). The distribution $p_{n_c}(\tau)$ at the critical point follows an algebraic decay $\sim \tau^{-9/8}$. For values $n<n_c$ in the ergodic phase, $p_n(\tau)$ bends downwards to a faster-than-algebraic decay. \textbf{c)} This faster decay is of the form of a streched exponential, as confirmed by inspecting $g_n(\tau)=p_n(\tau)/p_{n_c}(\tau) \sim \exp\{-(\tau/\tau_c)^{1/4}\}$. Plotting this quantity as a function of $\tau |n-n_c|^8$ leads to a scaling collapse at late times, consistent with $\tau_c \sim |n-n_c|^{-2z}=|n-n_c|^{-8}$. The densities shown are $n=1.2 - 1.4$ in steps of $0.05$.
}
\label{fig:3}
\end{figure*}

\textbf{\textit{Numerical analysis \& discussion.}}--
We can verify the predictions made in Eqs.~(\ref{eq:5}-\ref{eq:7}) in more detail by numerically evaluating the OTOC front $\overline{\braket{x_r(t)}}$ and the waiting time distribution $p_n(\tau)$, which is accessible through the single-shot OTOCs $C^{}_{ZX}(x,t;\bs{n})$. \figc{fig:3}{a} shows the propagation of the front $\overline{\braket{x_r(t)}}$ for different densities $n$.
For densities $n<n_c$ below the critical point, $\overline{\braket{x_r(t)}}$ first enters a sub-ballistic regime but starts to bend upwards at late times. Close to half-filling, the simulated times are sufficient to observe the bend continue until ballistic propagation is reached, \eq{eq:7}. The time at which the upwards bend sets in grows larger upon approaching the critical point. This is due to the algebraic contribution to the waiting time distribution that impacts the dynamics prior to the timescale on which the stretched exponential in \eq{eq:5} becomes fully relevant.
At the critical density $n=n_c$, no such bend is visible on the accessible timescales and our numerical findings follow the $\sim t^{1/2z} = t^{1/8}$ spread predicted in \eq{eq:7}. We further note that Eq.~(4) additionally predicts a $t^{1/2z}$--scaling of the \textit{broadening} of the OTOC front, which we have also verified numerically.
In the localized phase $n>n_c$, \figc{fig:3}{a} demonstrates that $\overline{\braket{x_r(t)}} < \mathrm{const.}$ remains bounded and the OTOC freezes at long times.

We show the numerically sampled distributions $p_n(\tau)$ in \figc{fig:3}{b} and indeed find a decay $\sim \tau^{-(1+1/2z)}=\tau^{-9/8}$ at the critical point. For $n<n_c$, the distribution $p_n(\tau)$ follows the critical line for some time and eventually crosses over into a faster-than-algebraic decay. We check that this faster decay corresponds to the stretched exponential of \eq{eq:5} by plotting the quantity $g_n(\tau):=p_n(\tau)/p_{n_c}(\tau)$, which we expect to go as $\exp\bigl\{ - (\tau/\tau_c)^{1/4} \bigr\}$ for long times; \figc{fig:3}{c}. 
Plotting $g_n(\tau)$ as a function of $\tau |n-n_c|^8$ leads to a scaling collapse at late times
, in agreement with \eq{eq:6}.

Having collected numerical evidence in favor of the phenomenologically derived distributions $p_n(\tau)$, we summarize our results via a dynamical phase diagram that captures the spread of the OTOC front in \figc{fig:1}{a}: In the long time limit, the spread is described by \eq{eq:7}. For finite times and below the critical density, we predict the OTOC to cross over from a sub-ballistic to a ballistic spread on a time scale given by the crossover time $\tau_c$ of the distribution $p_n(\tau)$ in \eq{eq:6}. We expect a similar crossover time scale to arise on the frozen side of the transition, as the longest but finite waiting times are of order $\xi^z \sim |n-n_c|^{2z}$, with $\xi$ the correlation length in the frozen phase~\cite{morningstar2020_kinetic}. We emphasize that due to the approximation of neglecting potentially parameter-dependent prefactors in the relation $\tau \sim \ell^z$, which we assumed in our phenomenological argument, the spreading of the OTOC at the critical point can in principle experience corrections. However, on our numerically accessible timescales, the Eqs.~(\ref{eq:5}-\ref{eq:7}) appear fully consistent.\\

\textbf{\textit{Conclusions \& outlook.}}--
We have analyzed the dynamics of operators in a dipole-conserving chain through the spreading of out-of-time-order correlators (OTOCs). The exotic fracton-like mobility constraints lead to an intricate phase diagram of operator dynamics, \figc{fig:1}{a}, that features a critical point where the OTOC front spreads \textit{sub-ballistically}. 
From the scaling properties of this critical point we derived a `critical fan', within which operators spread sub-ballistically at finite times even away from the critical density.
The slow operator spread $\sim t^{1/8}$ also bounds the growth of entanglement as well as charge transport, which is obstructed by the same localized regions. Our results suggest for the associated critical transport exponent $z_{\mathrm{tr}} \geq 8$, consistent with the numerical findings of Ref.~\cite{morningstar2020_kinetic}. The exact transport properties at the critical point are an interesting open question.

While we numerically studied specific automaton circuits, the two main ingredients of our phenomenological description, an ergodic dynamical phase with $z=4$ and a localized dynamical phase, can be found more generally for systems with dipole-conservation~\cite{Guardado20,feldmeier2020anomalous,Sala19,khemani20192d}. Scaling properties at the critical point should not be affected by specific model parameters. We therefore expect the same critical operator dynamics to apply universally to dipole-conserving chains.
Furthermore, while we focused on the $C^{(n)}_{ZX}$-OTOC, the sub-ballistic spread is expected to be robust to an arbitrary choice of local operators in \eq{eq:3}.
Specifically, we show in \cite{suppmat}, Sec. B, that for automaton circuits, the front of the $XX$-OTOC is bounded up to a factor of two by the front of the $ZX$-OTOC. 

We conjecture that our results also generalize to systems conserving all multipoles up to the $m$-th moment, where the above findings should be amended simply by substituting the subdiffusive transport exponents $z \rightarrow 2(m+1)$, derived in Refs.~\cite{gromov2020_fractonhydro,feldmeier2020anomalous}. Our results pave the way for finding new universality classes of operator growth, which might bear relevance to fractonic models in higher dimensions~\cite{feldmeier2021_fractondimer} and other constrained systems.\\

\textbf{\textit{Acknowledgements.}}-- We thank Giuseppe De Tomasi, David Huse, Alan Morningstar, Frank Pollmann and Pablo Sala for insightful discussions. We acknowledge support from the Technical University of Munich - Institute for Advanced Study, funded by the German Excellence Initiative and the European Union FP7 under grant agreement 291763, the Deutsche Forschungsgemeinschaft (DFG, German Research Foundation) under Germany’s Excellence Strategy-EXC-2111-390814868, TRR80 and DFG grant No. KN1254/2-1, No. KN1254/1-2, and from the European Research Council (ERC) under the European Union’s Horizon 2020 research and innovation programme (grant agreement No. 851161). We thank the Nanosystems Initiative Munich (NIM) funded by the German Excellence Initiative and the Leibniz Supercomputing Centre for access to their computational resources.

\bibliography{otoc}

\begin{thebibliography}{57}%
\makeatletter
\providecommand \@ifxundefined [1]{%
 \@ifx{#1\undefined}
}%
\providecommand \@ifnum [1]{%
 \ifnum #1\expandafter \@firstoftwo
 \else \expandafter \@secondoftwo
 \fi
}%
\providecommand \@ifx [1]{%
 \ifx #1\expandafter \@firstoftwo
 \else \expandafter \@secondoftwo
 \fi
}%
\providecommand \natexlab [1]{#1}%
\providecommand \enquote  [1]{``#1''}%
\providecommand \bibnamefont  [1]{#1}%
\providecommand \bibfnamefont [1]{#1}%
\providecommand \citenamefont [1]{#1}%
\providecommand \href@noop [0]{\@secondoftwo}%
\providecommand \href [0]{\begingroup \@sanitize@url \@href}%
\providecommand \@href[1]{\@@startlink{#1}\@@href}%
\providecommand \@@href[1]{\endgroup#1\@@endlink}%
\providecommand \@sanitize@url [0]{\catcode `\\12\catcode `\$12\catcode
  `\&12\catcode `\#12\catcode `\^12\catcode `\_12\catcode `\%12\relax}%
\providecommand \@@startlink[1]{}%
\providecommand \@@endlink[0]{}%
\providecommand \url  [0]{\begingroup\@sanitize@url \@url }%
\providecommand \@url [1]{\endgroup\@href {#1}{\urlprefix }}%
\providecommand \urlprefix  [0]{URL }%
\providecommand \Eprint [0]{\href }%
\providecommand \doibase [0]{https://doi.org/}%
\providecommand \selectlanguage [0]{\@gobble}%
\providecommand \bibinfo  [0]{\@secondoftwo}%
\providecommand \bibfield  [0]{\@secondoftwo}%
\providecommand \translation [1]{[#1]}%
\providecommand \BibitemOpen [0]{}%
\providecommand \bibitemStop [0]{}%
\providecommand \bibitemNoStop [0]{.\EOS\space}%
\providecommand \EOS [0]{\spacefactor3000\relax}%
\providecommand \BibitemShut  [1]{\csname bibitem#1\endcsname}%
\let\auto@bib@innerbib\@empty
\bibitem [{\citenamefont {Deutsch}(1991)}]{Deutsch91}%
  \BibitemOpen
  \bibfield  {author} {\bibinfo {author} {\bibfnamefont {J.~M.}\ \bibnamefont
  {Deutsch}},\ }\bibfield  {title} {\bibinfo {title} {{Quantum statistical
  mechanics in a closed system}},\ }\href
  {https://doi.org/10.1103/PhysRevA.43.2046} {\bibfield  {journal} {\bibinfo
  {journal} {Phys. Rev. A}\ }\textbf {\bibinfo {volume} {43}},\ \bibinfo
  {pages} {2046} (\bibinfo {year} {1991})}\BibitemShut {NoStop}%
\bibitem [{\citenamefont {Srednicki}(1994)}]{Srednicki94}%
  \BibitemOpen
  \bibfield  {author} {\bibinfo {author} {\bibfnamefont {M.}~\bibnamefont
  {Srednicki}},\ }\bibfield  {title} {\bibinfo {title} {{Chaos and quantum
  thermalization}},\ }\href {https://doi.org/10.1103/PhysRevE.50.888}
  {\bibfield  {journal} {\bibinfo  {journal} {Phys. Rev. E}\ }\textbf {\bibinfo
  {volume} {50}},\ \bibinfo {pages} {888} (\bibinfo {year} {1994})}\BibitemShut
  {NoStop}%
\bibitem [{\citenamefont {Rigol}\ \emph {et~al.}(2008)\citenamefont {Rigol},
  \citenamefont {Dunjko},\ and\ \citenamefont {Olshanii}}]{Rigol2008}%
  \BibitemOpen
  \bibfield  {author} {\bibinfo {author} {\bibfnamefont {M.}~\bibnamefont
  {Rigol}}, \bibinfo {author} {\bibfnamefont {V.}~\bibnamefont {Dunjko}},\ and\
  \bibinfo {author} {\bibfnamefont {M.}~\bibnamefont {Olshanii}},\ }\bibfield
  {title} {\bibinfo {title} {Thermalization and its mechanism for generic
  isolated quantum systems},\ }\href {https://doi.org/10.1038/nature06838}
  {\bibfield  {journal} {\bibinfo  {journal} {Nature}\ }\textbf {\bibinfo
  {volume} {452}},\ \bibinfo {pages} {854} (\bibinfo {year}
  {2008})}\BibitemShut {NoStop}%
\bibitem [{\citenamefont {D'Alessio}\ \emph {et~al.}(2016)\citenamefont
  {D'Alessio}, \citenamefont {Kafri}, \citenamefont {Polkovnikov},\ and\
  \citenamefont {Rigol}}]{alessio2016_chaos}%
  \BibitemOpen
  \bibfield  {author} {\bibinfo {author} {\bibfnamefont {L.}~\bibnamefont
  {D'Alessio}}, \bibinfo {author} {\bibfnamefont {Y.}~\bibnamefont {Kafri}},
  \bibinfo {author} {\bibfnamefont {A.}~\bibnamefont {Polkovnikov}},\ and\
  \bibinfo {author} {\bibfnamefont {M.}~\bibnamefont {Rigol}},\ }\bibfield
  {title} {\bibinfo {title} {{From quantum chaos and eigenstate thermalization
  to statistical mechanics and thermodynamics}},\ }\href
  {https://doi.org/10.1080/00018732.2016.1198134} {\bibfield  {journal}
  {\bibinfo  {journal} {Advances in Physics}\ }\textbf {\bibinfo {volume}
  {65}},\ \bibinfo {pages} {239} (\bibinfo {year} {2016})}\BibitemShut
  {NoStop}%
\bibitem [{\citenamefont {Kaufman}\ \emph {et~al.}(2016)\citenamefont
  {Kaufman}, \citenamefont {Tai}, \citenamefont {Lukin}, \citenamefont
  {Rispoli}, \citenamefont {Schittko}, \citenamefont {Preiss},\ and\
  \citenamefont {Greiner}}]{kaufman2016_thermalization}%
  \BibitemOpen
  \bibfield  {author} {\bibinfo {author} {\bibfnamefont {A.~M.}\ \bibnamefont
  {Kaufman}}, \bibinfo {author} {\bibfnamefont {M.~E.}\ \bibnamefont {Tai}},
  \bibinfo {author} {\bibfnamefont {A.}~\bibnamefont {Lukin}}, \bibinfo
  {author} {\bibfnamefont {M.}~\bibnamefont {Rispoli}}, \bibinfo {author}
  {\bibfnamefont {R.}~\bibnamefont {Schittko}}, \bibinfo {author}
  {\bibfnamefont {P.~M.}\ \bibnamefont {Preiss}},\ and\ \bibinfo {author}
  {\bibfnamefont {M.}~\bibnamefont {Greiner}},\ }\bibfield  {title} {\bibinfo
  {title} {{Quantum thermalization through entanglement in an isolated
  many-body system}},\ }\href {https://doi.org/10.1126/science.aaf6725}
  {\bibfield  {journal} {\bibinfo  {journal} {Science}\ }\textbf {\bibinfo
  {volume} {353}},\ \bibinfo {pages} {794} (\bibinfo {year}
  {2016})}\BibitemShut {NoStop}%
\bibitem [{\citenamefont {Brydges}\ \emph {et~al.}(2019)\citenamefont
  {Brydges}, \citenamefont {Elben}, \citenamefont {Jurcevic}, \citenamefont
  {Vermersch}, \citenamefont {Maier}, \citenamefont {Lanyon}, \citenamefont
  {Zoller}, \citenamefont {Blatt},\ and\ \citenamefont
  {Roos}}]{Brydges2019_renyi}%
  \BibitemOpen
  \bibfield  {author} {\bibinfo {author} {\bibfnamefont {T.}~\bibnamefont
  {Brydges}}, \bibinfo {author} {\bibfnamefont {A.}~\bibnamefont {Elben}},
  \bibinfo {author} {\bibfnamefont {P.}~\bibnamefont {Jurcevic}}, \bibinfo
  {author} {\bibfnamefont {B.}~\bibnamefont {Vermersch}}, \bibinfo {author}
  {\bibfnamefont {C.}~\bibnamefont {Maier}}, \bibinfo {author} {\bibfnamefont
  {B.~P.}\ \bibnamefont {Lanyon}}, \bibinfo {author} {\bibfnamefont
  {P.}~\bibnamefont {Zoller}}, \bibinfo {author} {\bibfnamefont
  {R.}~\bibnamefont {Blatt}},\ and\ \bibinfo {author} {\bibfnamefont {C.~F.}\
  \bibnamefont {Roos}},\ }\bibfield  {title} {\bibinfo {title} {{Probing
  R{\'e}nyi entanglement entropy via randomized measurements}},\ }\href
  {https://doi.org/10.1126/science.aau4963} {\bibfield  {journal} {\bibinfo
  {journal} {Science}\ }\textbf {\bibinfo {volume} {364}},\ \bibinfo {pages}
  {260} (\bibinfo {year} {2019})}\BibitemShut {NoStop}%
\bibitem [{\citenamefont {Chaikin}\ and\ \citenamefont
  {Lubensky}(1995)}]{chaikin_lubensky_1995}%
  \BibitemOpen
  \bibfield  {author} {\bibinfo {author} {\bibfnamefont {P.~M.}\ \bibnamefont
  {Chaikin}}\ and\ \bibinfo {author} {\bibfnamefont {T.~C.}\ \bibnamefont
  {Lubensky}},\ }\href {https://doi.org/10.1017/CBO9780511813467} {\emph
  {\bibinfo {title} {Principles of Condensed Matter Physics}}}\ (\bibinfo
  {publisher} {Cambridge University Press},\ \bibinfo {year}
  {1995})\BibitemShut {NoStop}%
\bibitem [{\citenamefont {Mukerjee}\ \emph {et~al.}(2006)\citenamefont
  {Mukerjee}, \citenamefont {Oganesyan},\ and\ \citenamefont
  {Huse}}]{Mukerjee06}%
  \BibitemOpen
  \bibfield  {author} {\bibinfo {author} {\bibfnamefont {S.}~\bibnamefont
  {Mukerjee}}, \bibinfo {author} {\bibfnamefont {V.}~\bibnamefont
  {Oganesyan}},\ and\ \bibinfo {author} {\bibfnamefont {D.}~\bibnamefont
  {Huse}},\ }\bibfield  {title} {\bibinfo {title} {{Statistical theory of
  transport by strongly interacting lattice fermions}},\ }\href
  {https://doi.org/10.1103/PhysRevB.73.035113} {\bibfield  {journal} {\bibinfo
  {journal} {Phys. Rev. B}\ }\textbf {\bibinfo {volume} {73}},\ \bibinfo
  {pages} {035113} (\bibinfo {year} {2006})}\BibitemShut {NoStop}%
\bibitem [{\citenamefont {Lux}\ \emph {et~al.}(2014)\citenamefont {Lux},
  \citenamefont {M\"uller}, \citenamefont {Mitra},\ and\ \citenamefont
  {Rosch}}]{Lux14}%
  \BibitemOpen
  \bibfield  {author} {\bibinfo {author} {\bibfnamefont {J.}~\bibnamefont
  {Lux}}, \bibinfo {author} {\bibfnamefont {J.}~\bibnamefont {M\"uller}},
  \bibinfo {author} {\bibfnamefont {A.}~\bibnamefont {Mitra}},\ and\ \bibinfo
  {author} {\bibfnamefont {A.}~\bibnamefont {Rosch}},\ }\bibfield  {title}
  {\bibinfo {title} {{Hydrodynamic long-time tails after a quantum quench}},\
  }\href {https://doi.org/10.1103/PhysRevA.89.053608} {\bibfield  {journal}
  {\bibinfo  {journal} {Phys. Rev. A}\ }\textbf {\bibinfo {volume} {89}},\
  \bibinfo {pages} {053608} (\bibinfo {year} {2014})}\BibitemShut {NoStop}%
\bibitem [{\citenamefont {Bohrdt}\ \emph {et~al.}(2017)\citenamefont {Bohrdt},
  \citenamefont {Mendl}, \citenamefont {Endres},\ and\ \citenamefont
  {Knap}}]{Bohrdt16}%
  \BibitemOpen
  \bibfield  {author} {\bibinfo {author} {\bibfnamefont {A.}~\bibnamefont
  {Bohrdt}}, \bibinfo {author} {\bibfnamefont {C.~B.}\ \bibnamefont {Mendl}},
  \bibinfo {author} {\bibfnamefont {M.}~\bibnamefont {Endres}},\ and\ \bibinfo
  {author} {\bibfnamefont {M.}~\bibnamefont {Knap}},\ }\bibfield  {title}
  {\bibinfo {title} {Scrambling and thermalization in a diffusive quantum
  many-body system},\ }\href {http://stacks.iop.org/1367-2630/19/i=6/a=063001}
  {\bibfield  {journal} {\bibinfo  {journal} {New Journal of Physics}\ }\textbf
  {\bibinfo {volume} {19}},\ \bibinfo {pages} {063001} (\bibinfo {year}
  {2017})}\BibitemShut {NoStop}%
\bibitem [{\citenamefont {Nahum}\ \emph {et~al.}(2017)\citenamefont {Nahum},
  \citenamefont {Ruhman}, \citenamefont {Vijay},\ and\ \citenamefont
  {Haah}}]{nahum2017_entanglement}%
  \BibitemOpen
  \bibfield  {author} {\bibinfo {author} {\bibfnamefont {A.}~\bibnamefont
  {Nahum}}, \bibinfo {author} {\bibfnamefont {J.}~\bibnamefont {Ruhman}},
  \bibinfo {author} {\bibfnamefont {S.}~\bibnamefont {Vijay}},\ and\ \bibinfo
  {author} {\bibfnamefont {J.}~\bibnamefont {Haah}},\ }\bibfield  {title}
  {\bibinfo {title} {{Quantum Entanglement Growth under Random Unitary
  Dynamics}},\ }\href {https://doi.org/10.1103/PhysRevX.7.031016} {\bibfield
  {journal} {\bibinfo  {journal} {Phys. Rev. X}\ }\textbf {\bibinfo {volume}
  {7}},\ \bibinfo {pages} {031016} (\bibinfo {year} {2017})}\BibitemShut
  {NoStop}%
\bibitem [{\citenamefont {Jonay}\ \emph {et~al.}(2018)\citenamefont {Jonay},
  \citenamefont {Huse},\ and\ \citenamefont {Nahum}}]{jonay2018_coarse}%
  \BibitemOpen
  \bibfield  {author} {\bibinfo {author} {\bibfnamefont {C.}~\bibnamefont
  {Jonay}}, \bibinfo {author} {\bibfnamefont {D.~A.}\ \bibnamefont {Huse}},\
  and\ \bibinfo {author} {\bibfnamefont {A.}~\bibnamefont {Nahum}},\
  }\href@noop {} {\bibinfo {title} {{Coarse-grained dynamics of operator and
  state entanglement}}} (\bibinfo {year} {2018}),\ \Eprint
  {https://arxiv.org/abs/1803.00089} {arXiv:1803.00089 [cond-mat.stat-mech]}
  \BibitemShut {NoStop}%
\bibitem [{\citenamefont {Knap}(2018)}]{knap2018_scrambling}%
  \BibitemOpen
  \bibfield  {author} {\bibinfo {author} {\bibfnamefont {M.}~\bibnamefont
  {Knap}},\ }\bibfield  {title} {\bibinfo {title} {{Entanglement production and
  information scrambling in a noisy spin system}},\ }\href
  {https://doi.org/10.1103/PhysRevB.98.184416} {\bibfield  {journal} {\bibinfo
  {journal} {Phys. Rev. B}\ }\textbf {\bibinfo {volume} {98}},\ \bibinfo
  {pages} {184416} (\bibinfo {year} {2018})}\BibitemShut {NoStop}%
\bibitem [{\citenamefont {Rakovszky}\ \emph
  {et~al.}(2019{\natexlab{a}})\citenamefont {Rakovszky}, \citenamefont
  {Pollmann},\ and\ \citenamefont {von Keyserlingk}}]{rakovszky2019_renyi}%
  \BibitemOpen
  \bibfield  {author} {\bibinfo {author} {\bibfnamefont {T.}~\bibnamefont
  {Rakovszky}}, \bibinfo {author} {\bibfnamefont {F.}~\bibnamefont
  {Pollmann}},\ and\ \bibinfo {author} {\bibfnamefont {C.~W.}\ \bibnamefont
  {von Keyserlingk}},\ }\bibfield  {title} {\bibinfo {title} {{Sub-ballistic
  Growth of R\'enyi Entropies due to Diffusion}},\ }\href
  {https://doi.org/10.1103/PhysRevLett.122.250602} {\bibfield  {journal}
  {\bibinfo  {journal} {Phys. Rev. Lett.}\ }\textbf {\bibinfo {volume} {122}},\
  \bibinfo {pages} {250602} (\bibinfo {year} {2019}{\natexlab{a}})}\BibitemShut
  {NoStop}%
\bibitem [{\citenamefont {Rakovszky}\ \emph
  {et~al.}(2019{\natexlab{b}})\citenamefont {Rakovszky}, \citenamefont {von
  Keyserlingk},\ and\ \citenamefont {Pollmann}}]{rakovszky2019_entanglement}%
  \BibitemOpen
  \bibfield  {author} {\bibinfo {author} {\bibfnamefont {T.}~\bibnamefont
  {Rakovszky}}, \bibinfo {author} {\bibfnamefont {C.~W.}\ \bibnamefont {von
  Keyserlingk}},\ and\ \bibinfo {author} {\bibfnamefont {F.}~\bibnamefont
  {Pollmann}},\ }\bibfield  {title} {\bibinfo {title} {{Entanglement growth
  after inhomogenous quenches}},\ }\href
  {https://doi.org/10.1103/PhysRevB.100.125139} {\bibfield  {journal} {\bibinfo
   {journal} {Phys. Rev. B}\ }\textbf {\bibinfo {volume} {100}},\ \bibinfo
  {pages} {125139} (\bibinfo {year} {2019}{\natexlab{b}})}\BibitemShut
  {NoStop}%
\bibitem [{\citenamefont {Nahum}\ \emph
  {et~al.}(2018{\natexlab{a}})\citenamefont {Nahum}, \citenamefont {Vijay},\
  and\ \citenamefont {Haah}}]{nahum2018_operator}%
  \BibitemOpen
  \bibfield  {author} {\bibinfo {author} {\bibfnamefont {A.}~\bibnamefont
  {Nahum}}, \bibinfo {author} {\bibfnamefont {S.}~\bibnamefont {Vijay}},\ and\
  \bibinfo {author} {\bibfnamefont {J.}~\bibnamefont {Haah}},\ }\bibfield
  {title} {\bibinfo {title} {{Operator Spreading in Random Unitary Circuits}},\
  }\href {https://doi.org/10.1103/PhysRevX.8.021014} {\bibfield  {journal}
  {\bibinfo  {journal} {Phys. Rev. X}\ }\textbf {\bibinfo {volume} {8}},\
  \bibinfo {pages} {021014} (\bibinfo {year} {2018}{\natexlab{a}})}\BibitemShut
  {NoStop}%
\bibitem [{\citenamefont {von Keyserlingk}\ \emph {et~al.}(2018)\citenamefont
  {von Keyserlingk}, \citenamefont {Rakovszky}, \citenamefont {Pollmann},\ and\
  \citenamefont {Sondhi}}]{Keyserlingk2018}%
  \BibitemOpen
  \bibfield  {author} {\bibinfo {author} {\bibfnamefont {C.~W.}\ \bibnamefont
  {von Keyserlingk}}, \bibinfo {author} {\bibfnamefont {T.}~\bibnamefont
  {Rakovszky}}, \bibinfo {author} {\bibfnamefont {F.}~\bibnamefont
  {Pollmann}},\ and\ \bibinfo {author} {\bibfnamefont {S.~L.}\ \bibnamefont
  {Sondhi}},\ }\bibfield  {title} {\bibinfo {title} {{Operator Hydrodynamics,
  OTOCs, and Entanglement Growth in Systems without Conservation Laws}},\
  }\href {https://doi.org/10.1103/PhysRevX.8.021013} {\bibfield  {journal}
  {\bibinfo  {journal} {Phys. Rev. X}\ }\textbf {\bibinfo {volume} {8}},\
  \bibinfo {pages} {021013} (\bibinfo {year} {2018})}\BibitemShut {NoStop}%
\bibitem [{\citenamefont {Khemani}\ \emph {et~al.}(2018)\citenamefont
  {Khemani}, \citenamefont {Vishwanath},\ and\ \citenamefont
  {Huse}}]{khemani2018_operator}%
  \BibitemOpen
  \bibfield  {author} {\bibinfo {author} {\bibfnamefont {V.}~\bibnamefont
  {Khemani}}, \bibinfo {author} {\bibfnamefont {A.}~\bibnamefont
  {Vishwanath}},\ and\ \bibinfo {author} {\bibfnamefont {D.~A.}\ \bibnamefont
  {Huse}},\ }\bibfield  {title} {\bibinfo {title} {{Operator Spreading and the
  Emergence of Dissipative Hydrodynamics under Unitary Evolution with
  Conservation Laws}},\ }\href {https://doi.org/10.1103/PhysRevX.8.031057}
  {\bibfield  {journal} {\bibinfo  {journal} {Phys. Rev. X}\ }\textbf {\bibinfo
  {volume} {8}},\ \bibinfo {pages} {031057} (\bibinfo {year}
  {2018})}\BibitemShut {NoStop}%
\bibitem [{\citenamefont {Rakovszky}\ \emph {et~al.}(2018)\citenamefont
  {Rakovszky}, \citenamefont {Pollmann},\ and\ \citenamefont {von
  Keyserlingk}}]{Rakovszky18}%
  \BibitemOpen
  \bibfield  {author} {\bibinfo {author} {\bibfnamefont {T.}~\bibnamefont
  {Rakovszky}}, \bibinfo {author} {\bibfnamefont {F.}~\bibnamefont
  {Pollmann}},\ and\ \bibinfo {author} {\bibfnamefont {C.~W.}\ \bibnamefont
  {von Keyserlingk}},\ }\bibfield  {title} {\bibinfo {title} {{Diffusive
  Hydrodynamics of Out-of-Time-Ordered Correlators with Charge Conservation}},\
  }\href {https://doi.org/10.1103/PhysRevX.8.031058} {\bibfield  {journal}
  {\bibinfo  {journal} {Phys. Rev. X}\ }\textbf {\bibinfo {volume} {8}},\
  \bibinfo {pages} {031058} (\bibinfo {year} {2018})}\BibitemShut {NoStop}%
\bibitem [{\citenamefont {Nahum}\ \emph
  {et~al.}(2018{\natexlab{b}})\citenamefont {Nahum}, \citenamefont {Ruhman},\
  and\ \citenamefont {Huse}}]{nahum2018_griffith}%
  \BibitemOpen
  \bibfield  {author} {\bibinfo {author} {\bibfnamefont {A.}~\bibnamefont
  {Nahum}}, \bibinfo {author} {\bibfnamefont {J.}~\bibnamefont {Ruhman}},\ and\
  \bibinfo {author} {\bibfnamefont {D.~A.}\ \bibnamefont {Huse}},\ }\bibfield
  {title} {\bibinfo {title} {{Dynamics of entanglement and transport in
  one-dimensional systems with quenched randomness}},\ }\href
  {https://doi.org/10.1103/PhysRevB.98.035118} {\bibfield  {journal} {\bibinfo
  {journal} {Phys. Rev. B}\ }\textbf {\bibinfo {volume} {98}},\ \bibinfo
  {pages} {035118} (\bibinfo {year} {2018}{\natexlab{b}})}\BibitemShut
  {NoStop}%
\bibitem [{\citenamefont {Parker}\ \emph {et~al.}(2019)\citenamefont {Parker},
  \citenamefont {Cao}, \citenamefont {Avdoshkin}, \citenamefont {Scaffidi},\
  and\ \citenamefont {Altman}}]{Parker19}%
  \BibitemOpen
  \bibfield  {author} {\bibinfo {author} {\bibfnamefont {D.~E.}\ \bibnamefont
  {Parker}}, \bibinfo {author} {\bibfnamefont {X.}~\bibnamefont {Cao}},
  \bibinfo {author} {\bibfnamefont {A.}~\bibnamefont {Avdoshkin}}, \bibinfo
  {author} {\bibfnamefont {T.}~\bibnamefont {Scaffidi}},\ and\ \bibinfo
  {author} {\bibfnamefont {E.}~\bibnamefont {Altman}},\ }\bibfield  {title}
  {\bibinfo {title} {{A Universal Operator Growth Hypothesis}},\ }\href
  {https://doi.org/10.1103/PhysRevX.9.041017} {\bibfield  {journal} {\bibinfo
  {journal} {Phys. Rev. X}\ }\textbf {\bibinfo {volume} {9}},\ \bibinfo {pages}
  {041017} (\bibinfo {year} {2019})}\BibitemShut {NoStop}%
\bibitem [{\citenamefont {Nandkishore}\ and\ \citenamefont
  {Hermele}(2019)}]{nandkishore2019_fractons}%
  \BibitemOpen
  \bibfield  {author} {\bibinfo {author} {\bibfnamefont {R.~M.}\ \bibnamefont
  {Nandkishore}}\ and\ \bibinfo {author} {\bibfnamefont {M.}~\bibnamefont
  {Hermele}},\ }\bibfield  {title} {\bibinfo {title} {Fractons},\ }\href
  {https://doi.org/10.1146/annurev-conmatphys-031218-013604} {\bibfield
  {journal} {\bibinfo  {journal} {Annual Review of Condensed Matter Physics}\
  }\textbf {\bibinfo {volume} {10}},\ \bibinfo {pages} {295} (\bibinfo {year}
  {2019})}\BibitemShut {NoStop}%
\bibitem [{\citenamefont {Pretko}\ \emph {et~al.}(2020)\citenamefont {Pretko},
  \citenamefont {Chen},\ and\ \citenamefont {You}}]{pretko2020_fracton}%
  \BibitemOpen
  \bibfield  {author} {\bibinfo {author} {\bibfnamefont {M.}~\bibnamefont
  {Pretko}}, \bibinfo {author} {\bibfnamefont {X.}~\bibnamefont {Chen}},\ and\
  \bibinfo {author} {\bibfnamefont {Y.}~\bibnamefont {You}},\ }\bibfield
  {title} {\bibinfo {title} {Fracton phases of matter},\ }\href@noop {}
  {\bibfield  {journal} {\bibinfo  {journal} {International Journal of Modern
  Physics A}\ }\textbf {\bibinfo {volume} {35}},\ \bibinfo {pages} {2030003}
  (\bibinfo {year} {2020})}\BibitemShut {NoStop}%
\bibitem [{\citenamefont {Chamon}(2005)}]{chamon2005_glass}%
  \BibitemOpen
  \bibfield  {author} {\bibinfo {author} {\bibfnamefont {C.}~\bibnamefont
  {Chamon}},\ }\bibfield  {title} {\bibinfo {title} {{Quantum Glassiness in
  Strongly Correlated Clean Systems: An Example of Topological
  Overprotection}},\ }\href {https://doi.org/10.1103/PhysRevLett.94.040402}
  {\bibfield  {journal} {\bibinfo  {journal} {Phys. Rev. Lett.}\ }\textbf
  {\bibinfo {volume} {94}},\ \bibinfo {pages} {040402} (\bibinfo {year}
  {2005})}\BibitemShut {NoStop}%
\bibitem [{\citenamefont {Haah}(2011)}]{haah2011_code}%
  \BibitemOpen
  \bibfield  {author} {\bibinfo {author} {\bibfnamefont {J.}~\bibnamefont
  {Haah}},\ }\bibfield  {title} {\bibinfo {title} {{Local stabilizer codes in
  three dimensions without string logical operators}},\ }\href
  {https://doi.org/10.1103/PhysRevA.83.042330} {\bibfield  {journal} {\bibinfo
  {journal} {Phys. Rev. A}\ }\textbf {\bibinfo {volume} {83}},\ \bibinfo
  {pages} {042330} (\bibinfo {year} {2011})}\BibitemShut {NoStop}%
\bibitem [{\citenamefont {Yoshida}(2013)}]{yoshida2013_fractal}%
  \BibitemOpen
  \bibfield  {author} {\bibinfo {author} {\bibfnamefont {B.}~\bibnamefont
  {Yoshida}},\ }\bibfield  {title} {\bibinfo {title} {{Exotic topological order
  in fractal spin liquids}},\ }\href
  {https://doi.org/10.1103/PhysRevB.88.125122} {\bibfield  {journal} {\bibinfo
  {journal} {Phys. Rev. B}\ }\textbf {\bibinfo {volume} {88}},\ \bibinfo
  {pages} {125122} (\bibinfo {year} {2013})}\BibitemShut {NoStop}%
\bibitem [{\citenamefont {Vijay}\ \emph {et~al.}(2015)\citenamefont {Vijay},
  \citenamefont {Haah},\ and\ \citenamefont {Fu}}]{vijay2015_topo}%
  \BibitemOpen
  \bibfield  {author} {\bibinfo {author} {\bibfnamefont {S.}~\bibnamefont
  {Vijay}}, \bibinfo {author} {\bibfnamefont {J.}~\bibnamefont {Haah}},\ and\
  \bibinfo {author} {\bibfnamefont {L.}~\bibnamefont {Fu}},\ }\bibfield
  {title} {\bibinfo {title} {{A new kind of topological quantum order: A
  dimensional hierarchy of quasiparticles built from stationary excitations}},\
  }\href {https://doi.org/10.1103/PhysRevB.92.235136} {\bibfield  {journal}
  {\bibinfo  {journal} {Phys. Rev. B}\ }\textbf {\bibinfo {volume} {92}},\
  \bibinfo {pages} {235136} (\bibinfo {year} {2015})}\BibitemShut {NoStop}%
\bibitem [{\citenamefont {Vijay}\ \emph {et~al.}(2016)\citenamefont {Vijay},
  \citenamefont {Haah},\ and\ \citenamefont {Fu}}]{Vijay16}%
  \BibitemOpen
  \bibfield  {author} {\bibinfo {author} {\bibfnamefont {S.}~\bibnamefont
  {Vijay}}, \bibinfo {author} {\bibfnamefont {J.}~\bibnamefont {Haah}},\ and\
  \bibinfo {author} {\bibfnamefont {L.}~\bibnamefont {Fu}},\ }\bibfield
  {title} {\bibinfo {title} {Fracton topological order, generalized lattice
  gauge theory, and duality},\ }\href
  {https://doi.org/10.1103/PhysRevB.94.235157} {\bibfield  {journal} {\bibinfo
  {journal} {Phys. Rev. B}\ }\textbf {\bibinfo {volume} {94}},\ \bibinfo
  {pages} {235157} (\bibinfo {year} {2016})}\BibitemShut {NoStop}%
\bibitem [{\citenamefont {Pretko}\ and\ \citenamefont
  {Radzihovsky}(2018)}]{pretko2018_elasticity}%
  \BibitemOpen
  \bibfield  {author} {\bibinfo {author} {\bibfnamefont {M.}~\bibnamefont
  {Pretko}}\ and\ \bibinfo {author} {\bibfnamefont {L.}~\bibnamefont
  {Radzihovsky}},\ }\bibfield  {title} {\bibinfo {title} {{Fracton-Elasticity
  Duality}},\ }\href {https://doi.org/10.1103/PhysRevLett.120.195301}
  {\bibfield  {journal} {\bibinfo  {journal} {Phys. Rev. Lett.}\ }\textbf
  {\bibinfo {volume} {120}},\ \bibinfo {pages} {195301} (\bibinfo {year}
  {2018})}\BibitemShut {NoStop}%
\bibitem [{\citenamefont {Pretko}(2017{\natexlab{a}})}]{pretko2017_subdim}%
  \BibitemOpen
  \bibfield  {author} {\bibinfo {author} {\bibfnamefont {M.}~\bibnamefont
  {Pretko}},\ }\bibfield  {title} {\bibinfo {title} {{Subdimensional particle
  structure of higher rank $U(1)$ spin liquids}},\ }\href
  {https://doi.org/10.1103/PhysRevB.95.115139} {\bibfield  {journal} {\bibinfo
  {journal} {Phys. Rev. B}\ }\textbf {\bibinfo {volume} {95}},\ \bibinfo
  {pages} {115139} (\bibinfo {year} {2017}{\natexlab{a}})}\BibitemShut
  {NoStop}%
\bibitem [{\citenamefont {Pretko}(2018)}]{pretko2018_gaugprinciple}%
  \BibitemOpen
  \bibfield  {author} {\bibinfo {author} {\bibfnamefont {M.}~\bibnamefont
  {Pretko}},\ }\bibfield  {title} {\bibinfo {title} {The fracton gauge
  principle},\ }\href {https://doi.org/10.1103/PhysRevB.98.115134} {\bibfield
  {journal} {\bibinfo  {journal} {Phys. Rev. B}\ }\textbf {\bibinfo {volume}
  {98}},\ \bibinfo {pages} {115134} (\bibinfo {year} {2018})}\BibitemShut
  {NoStop}%
\bibitem [{\citenamefont {Pretko}(2017{\natexlab{b}})}]{pretko_witten}%
  \BibitemOpen
  \bibfield  {author} {\bibinfo {author} {\bibfnamefont {M.}~\bibnamefont
  {Pretko}},\ }\bibfield  {title} {\bibinfo {title} {{Higher-spin Witten effect
  and two-dimensional fracton phases}},\ }\href
  {https://doi.org/10.1103/PhysRevB.96.125151} {\bibfield  {journal} {\bibinfo
  {journal} {Phys. Rev. B}\ }\textbf {\bibinfo {volume} {96}},\ \bibinfo
  {pages} {125151} (\bibinfo {year} {2017}{\natexlab{b}})}\BibitemShut
  {NoStop}%
\bibitem [{\citenamefont {Williamson}\ \emph {et~al.}(2019)\citenamefont
  {Williamson}, \citenamefont {Bi},\ and\ \citenamefont
  {Cheng}}]{williamson2019_fractonic}%
  \BibitemOpen
  \bibfield  {author} {\bibinfo {author} {\bibfnamefont {D.~J.}\ \bibnamefont
  {Williamson}}, \bibinfo {author} {\bibfnamefont {Z.}~\bibnamefont {Bi}},\
  and\ \bibinfo {author} {\bibfnamefont {M.}~\bibnamefont {Cheng}},\ }\bibfield
   {title} {\bibinfo {title} {{Fractonic matter in symmetry-enriched $U(1)$
  gauge theory}},\ }\href {https://doi.org/10.1103/PhysRevB.100.125150}
  {\bibfield  {journal} {\bibinfo  {journal} {Phys. Rev. B}\ }\textbf {\bibinfo
  {volume} {100}},\ \bibinfo {pages} {125150} (\bibinfo {year}
  {2019})}\BibitemShut {NoStop}%
\bibitem [{\citenamefont {Sala}\ \emph {et~al.}(2020)\citenamefont {Sala},
  \citenamefont {Rakovszky}, \citenamefont {Verresen}, \citenamefont {Knap},\
  and\ \citenamefont {Pollmann}}]{Sala19}%
  \BibitemOpen
  \bibfield  {author} {\bibinfo {author} {\bibfnamefont {P.}~\bibnamefont
  {Sala}}, \bibinfo {author} {\bibfnamefont {T.}~\bibnamefont {Rakovszky}},
  \bibinfo {author} {\bibfnamefont {R.}~\bibnamefont {Verresen}}, \bibinfo
  {author} {\bibfnamefont {M.}~\bibnamefont {Knap}},\ and\ \bibinfo {author}
  {\bibfnamefont {F.}~\bibnamefont {Pollmann}},\ }\bibfield  {title} {\bibinfo
  {title} {{Ergodicity Breaking Arising from Hilbert Space Fragmentation in
  Dipole-Conserving Hamiltonians}},\ }\href
  {https://doi.org/10.1103/PhysRevX.10.011047} {\bibfield  {journal} {\bibinfo
  {journal} {Phys. Rev. X}\ }\textbf {\bibinfo {volume} {10}},\ \bibinfo
  {pages} {011047} (\bibinfo {year} {2020})}\BibitemShut {NoStop}%
\bibitem [{\citenamefont {Khemani}\ \emph {et~al.}(2020)\citenamefont
  {Khemani}, \citenamefont {Hermele},\ and\ \citenamefont
  {Nandkishore}}]{khemani20192d}%
  \BibitemOpen
  \bibfield  {author} {\bibinfo {author} {\bibfnamefont {V.}~\bibnamefont
  {Khemani}}, \bibinfo {author} {\bibfnamefont {M.}~\bibnamefont {Hermele}},\
  and\ \bibinfo {author} {\bibfnamefont {R.}~\bibnamefont {Nandkishore}},\
  }\bibfield  {title} {\bibinfo {title} {{Localization from Hilbert space
  shattering: From theory to physical realizations}},\ }\href
  {https://doi.org/10.1103/PhysRevB.101.174204} {\bibfield  {journal} {\bibinfo
   {journal} {Phys. Rev. B}\ }\textbf {\bibinfo {volume} {101}},\ \bibinfo
  {pages} {174204} (\bibinfo {year} {2020})}\BibitemShut {NoStop}%
\bibitem [{\citenamefont {Rakovszky}\ \emph {et~al.}(2020)\citenamefont
  {Rakovszky}, \citenamefont {Sala}, \citenamefont {Verresen}, \citenamefont
  {Knap},\ and\ \citenamefont {Pollmann}}]{Rakovszky20}%
  \BibitemOpen
  \bibfield  {author} {\bibinfo {author} {\bibfnamefont {T.}~\bibnamefont
  {Rakovszky}}, \bibinfo {author} {\bibfnamefont {P.}~\bibnamefont {Sala}},
  \bibinfo {author} {\bibfnamefont {R.}~\bibnamefont {Verresen}}, \bibinfo
  {author} {\bibfnamefont {M.}~\bibnamefont {Knap}},\ and\ \bibinfo {author}
  {\bibfnamefont {F.}~\bibnamefont {Pollmann}},\ }\bibfield  {title} {\bibinfo
  {title} {{Statistical localization: From strong fragmentation to strong edge
  modes}},\ }\href {https://doi.org/10.1103/PhysRevB.101.125126} {\bibfield
  {journal} {\bibinfo  {journal} {Phys. Rev. B}\ }\textbf {\bibinfo {volume}
  {101}},\ \bibinfo {pages} {125126} (\bibinfo {year} {2020})}\BibitemShut
  {NoStop}%
\bibitem [{\citenamefont {Scherg}\ \emph {et~al.}(2021)\citenamefont {Scherg},
  \citenamefont {Kohlert}, \citenamefont {Sala}, \citenamefont {Pollmann},
  \citenamefont {Hebbe~Madhusudhana}, \citenamefont {Bloch},\ and\
  \citenamefont {Aidelsburger}}]{scherg2021_kinetic}%
  \BibitemOpen
  \bibfield  {author} {\bibinfo {author} {\bibfnamefont {S.}~\bibnamefont
  {Scherg}}, \bibinfo {author} {\bibfnamefont {T.}~\bibnamefont {Kohlert}},
  \bibinfo {author} {\bibfnamefont {P.}~\bibnamefont {Sala}}, \bibinfo {author}
  {\bibfnamefont {F.}~\bibnamefont {Pollmann}}, \bibinfo {author}
  {\bibfnamefont {B.}~\bibnamefont {Hebbe~Madhusudhana}}, \bibinfo {author}
  {\bibfnamefont {I.}~\bibnamefont {Bloch}},\ and\ \bibinfo {author}
  {\bibfnamefont {M.}~\bibnamefont {Aidelsburger}},\ }\bibfield  {title}
  {\bibinfo {title} {{Observing non-ergodicity due to kinetic constraints in
  tilted Fermi-Hubbard chains}},\ }\href
  {https://doi.org/10.1038/s41467-021-24726-0} {\bibfield  {journal} {\bibinfo
  {journal} {Nature Communications}\ }\textbf {\bibinfo {volume} {12}},\
  \bibinfo {pages} {4490} (\bibinfo {year} {2021})}\BibitemShut {NoStop}%
\bibitem [{\citenamefont {Gromov}\ \emph {et~al.}(2020)\citenamefont {Gromov},
  \citenamefont {Lucas},\ and\ \citenamefont
  {Nandkishore}}]{gromov2020_fractonhydro}%
  \BibitemOpen
  \bibfield  {author} {\bibinfo {author} {\bibfnamefont {A.}~\bibnamefont
  {Gromov}}, \bibinfo {author} {\bibfnamefont {A.}~\bibnamefont {Lucas}},\ and\
  \bibinfo {author} {\bibfnamefont {R.~M.}\ \bibnamefont {Nandkishore}},\
  }\bibfield  {title} {\bibinfo {title} {Fracton hydrodynamics},\ }\href
  {https://doi.org/10.1103/PhysRevResearch.2.033124} {\bibfield  {journal}
  {\bibinfo  {journal} {Phys. Rev. Research}\ }\textbf {\bibinfo {volume}
  {2}},\ \bibinfo {pages} {033124} (\bibinfo {year} {2020})}\BibitemShut
  {NoStop}%
\bibitem [{\citenamefont {Feldmeier}\ \emph {et~al.}(2020)\citenamefont
  {Feldmeier}, \citenamefont {Sala}, \citenamefont {De~Tomasi}, \citenamefont
  {Pollmann},\ and\ \citenamefont {Knap}}]{feldmeier2020anomalous}%
  \BibitemOpen
  \bibfield  {author} {\bibinfo {author} {\bibfnamefont {J.}~\bibnamefont
  {Feldmeier}}, \bibinfo {author} {\bibfnamefont {P.}~\bibnamefont {Sala}},
  \bibinfo {author} {\bibfnamefont {G.}~\bibnamefont {De~Tomasi}}, \bibinfo
  {author} {\bibfnamefont {F.}~\bibnamefont {Pollmann}},\ and\ \bibinfo
  {author} {\bibfnamefont {M.}~\bibnamefont {Knap}},\ }\bibfield  {title}
  {\bibinfo {title} {{Anomalous Diffusion in Dipole- and
  Higher-Moment-Conserving Systems}},\ }\href
  {https://doi.org/10.1103/PhysRevLett.125.245303} {\bibfield  {journal}
  {\bibinfo  {journal} {Phys. Rev. Lett.}\ }\textbf {\bibinfo {volume} {125}},\
  \bibinfo {pages} {245303} (\bibinfo {year} {2020})}\BibitemShut {NoStop}%
\bibitem [{\citenamefont {Morningstar}\ \emph {et~al.}(2020)\citenamefont
  {Morningstar}, \citenamefont {Khemani},\ and\ \citenamefont
  {Huse}}]{morningstar2020_kinetic}%
  \BibitemOpen
  \bibfield  {author} {\bibinfo {author} {\bibfnamefont {A.}~\bibnamefont
  {Morningstar}}, \bibinfo {author} {\bibfnamefont {V.}~\bibnamefont
  {Khemani}},\ and\ \bibinfo {author} {\bibfnamefont {D.~A.}\ \bibnamefont
  {Huse}},\ }\bibfield  {title} {\bibinfo {title} {{Kinetically constrained
  freezing transition in a dipole-conserving system}},\ }\href
  {https://doi.org/10.1103/PhysRevB.101.214205} {\bibfield  {journal} {\bibinfo
   {journal} {Phys. Rev. B}\ }\textbf {\bibinfo {volume} {101}},\ \bibinfo
  {pages} {214205} (\bibinfo {year} {2020})}\BibitemShut {NoStop}%
\bibitem [{\citenamefont {Zhang}(2020)}]{zhang_2020}%
  \BibitemOpen
  \bibfield  {author} {\bibinfo {author} {\bibfnamefont {P.}~\bibnamefont
  {Zhang}},\ }\bibfield  {title} {\bibinfo {title} {Subdiffusion in strongly
  tilted lattice systems},\ }\href
  {https://doi.org/10.1103/PhysRevResearch.2.033129} {\bibfield  {journal}
  {\bibinfo  {journal} {Phys. Rev. Research}\ }\textbf {\bibinfo {volume}
  {2}},\ \bibinfo {pages} {033129} (\bibinfo {year} {2020})}\BibitemShut
  {NoStop}%
\bibitem [{\citenamefont {Moudgalya}\ \emph {et~al.}(2021)\citenamefont
  {Moudgalya}, \citenamefont {Prem}, \citenamefont {Huse},\ and\ \citenamefont
  {Chan}}]{moudgalya2021_spectral}%
  \BibitemOpen
  \bibfield  {author} {\bibinfo {author} {\bibfnamefont {S.}~\bibnamefont
  {Moudgalya}}, \bibinfo {author} {\bibfnamefont {A.}~\bibnamefont {Prem}},
  \bibinfo {author} {\bibfnamefont {D.~A.}\ \bibnamefont {Huse}},\ and\
  \bibinfo {author} {\bibfnamefont {A.}~\bibnamefont {Chan}},\ }\bibfield
  {title} {\bibinfo {title} {{Spectral statistics in constrained many-body
  quantum chaotic systems}},\ }\href
  {https://doi.org/10.1103/PhysRevResearch.3.023176} {\bibfield  {journal}
  {\bibinfo  {journal} {Phys. Rev. Research}\ }\textbf {\bibinfo {volume}
  {3}},\ \bibinfo {pages} {023176} (\bibinfo {year} {2021})}\BibitemShut
  {NoStop}%
\bibitem [{\citenamefont {Guardado-Sanchez}\ \emph {et~al.}(2020)\citenamefont
  {Guardado-Sanchez}, \citenamefont {Morningstar}, \citenamefont {Spar},
  \citenamefont {Brown}, \citenamefont {Huse},\ and\ \citenamefont
  {Bakr}}]{Guardado20}%
  \BibitemOpen
  \bibfield  {author} {\bibinfo {author} {\bibfnamefont {E.}~\bibnamefont
  {Guardado-Sanchez}}, \bibinfo {author} {\bibfnamefont {A.}~\bibnamefont
  {Morningstar}}, \bibinfo {author} {\bibfnamefont {B.~M.}\ \bibnamefont
  {Spar}}, \bibinfo {author} {\bibfnamefont {P.~T.}\ \bibnamefont {Brown}},
  \bibinfo {author} {\bibfnamefont {D.~A.}\ \bibnamefont {Huse}},\ and\
  \bibinfo {author} {\bibfnamefont {W.~S.}\ \bibnamefont {Bakr}},\ }\bibfield
  {title} {\bibinfo {title} {{Subdiffusion and Heat Transport in a Tilted
  Two-Dimensional Fermi-Hubbard System}},\ }\href
  {https://doi.org/10.1103/PhysRevX.10.011042} {\bibfield  {journal} {\bibinfo
  {journal} {Phys. Rev. X}\ }\textbf {\bibinfo {volume} {10}},\ \bibinfo
  {pages} {011042} (\bibinfo {year} {2020})}\BibitemShut {NoStop}%
\bibitem [{\citenamefont {Gopalakrishnan}(2018)}]{gopalakrishnan2018_operator}%
  \BibitemOpen
  \bibfield  {author} {\bibinfo {author} {\bibfnamefont {S.}~\bibnamefont
  {Gopalakrishnan}},\ }\bibfield  {title} {\bibinfo {title} {{Operator growth
  and eigenstate entanglement in an interacting integrable Floquet system}},\
  }\href {https://doi.org/10.1103/PhysRevB.98.060302} {\bibfield  {journal}
  {\bibinfo  {journal} {Phys. Rev. B}\ }\textbf {\bibinfo {volume} {98}},\
  \bibinfo {pages} {060302} (\bibinfo {year} {2018})}\BibitemShut {NoStop}%
\bibitem [{\citenamefont {Gopalakrishnan}\ and\ \citenamefont
  {Zakirov}(2018)}]{Gopalakrishnan_2018}%
  \BibitemOpen
  \bibfield  {author} {\bibinfo {author} {\bibfnamefont {S.}~\bibnamefont
  {Gopalakrishnan}}\ and\ \bibinfo {author} {\bibfnamefont {B.}~\bibnamefont
  {Zakirov}},\ }\bibfield  {title} {\bibinfo {title} {{Facilitated quantum
  cellular automata as simple models with non-thermal eigenstates and
  dynamics}},\ }\href {https://doi.org/10.1088/2058-9565/aad759} {\bibfield
  {journal} {\bibinfo  {journal} {Quantum Science and Technology}\ }\textbf
  {\bibinfo {volume} {3}},\ \bibinfo {pages} {044004} (\bibinfo {year}
  {2018})}\BibitemShut {NoStop}%
\bibitem [{\citenamefont {Iaconis}\ \emph {et~al.}(2019)\citenamefont
  {Iaconis}, \citenamefont {Vijay},\ and\ \citenamefont
  {Nandkishore}}]{Iaconis19}%
  \BibitemOpen
  \bibfield  {author} {\bibinfo {author} {\bibfnamefont {J.}~\bibnamefont
  {Iaconis}}, \bibinfo {author} {\bibfnamefont {S.}~\bibnamefont {Vijay}},\
  and\ \bibinfo {author} {\bibfnamefont {R.}~\bibnamefont {Nandkishore}},\
  }\bibfield  {title} {\bibinfo {title} {{Anomalous subdiffusion from subsystem
  symmetries}},\ }\href {https://doi.org/10.1103/PhysRevB.100.214301}
  {\bibfield  {journal} {\bibinfo  {journal} {Phys. Rev. B}\ }\textbf {\bibinfo
  {volume} {100}},\ \bibinfo {pages} {214301} (\bibinfo {year}
  {2019})}\BibitemShut {NoStop}%
\bibitem [{\citenamefont {Iaconis}(2021)}]{iaconis2021_complexity}%
  \BibitemOpen
  \bibfield  {author} {\bibinfo {author} {\bibfnamefont {J.}~\bibnamefont
  {Iaconis}},\ }\bibfield  {title} {\bibinfo {title} {{Quantum State Complexity
  in Computationally Tractable Quantum Circuits}},\ }\href
  {https://doi.org/10.1103/PRXQuantum.2.010329} {\bibfield  {journal} {\bibinfo
   {journal} {PRX Quantum}\ }\textbf {\bibinfo {volume} {2}},\ \bibinfo {pages}
  {010329} (\bibinfo {year} {2021})}\BibitemShut {NoStop}%
\bibitem [{\citenamefont {Iaconis}\ \emph {et~al.}(2020)\citenamefont
  {Iaconis}, \citenamefont {Lucas},\ and\ \citenamefont
  {Chen}}]{iaconis2020_measurement}%
  \BibitemOpen
  \bibfield  {author} {\bibinfo {author} {\bibfnamefont {J.}~\bibnamefont
  {Iaconis}}, \bibinfo {author} {\bibfnamefont {A.}~\bibnamefont {Lucas}},\
  and\ \bibinfo {author} {\bibfnamefont {X.}~\bibnamefont {Chen}},\ }\bibfield
  {title} {\bibinfo {title} {{Measurement-induced phase transitions in quantum
  automaton circuits}},\ }\href {https://doi.org/10.1103/PhysRevB.102.224311}
  {\bibfield  {journal} {\bibinfo  {journal} {Phys. Rev. B}\ }\textbf {\bibinfo
  {volume} {102}},\ \bibinfo {pages} {224311} (\bibinfo {year}
  {2020})}\BibitemShut {NoStop}%
\bibitem [{\citenamefont {Iaconis}\ \emph {et~al.}(2021)\citenamefont
  {Iaconis}, \citenamefont {Lucas},\ and\ \citenamefont
  {Nandkishore}}]{iaconis2021_multipole}%
  \BibitemOpen
  \bibfield  {author} {\bibinfo {author} {\bibfnamefont {J.}~\bibnamefont
  {Iaconis}}, \bibinfo {author} {\bibfnamefont {A.}~\bibnamefont {Lucas}},\
  and\ \bibinfo {author} {\bibfnamefont {R.}~\bibnamefont {Nandkishore}},\
  }\bibfield  {title} {\bibinfo {title} {{Multipole conservation laws and
  subdiffusion in any dimension}},\ }\href
  {https://doi.org/10.1103/PhysRevE.103.022142} {\bibfield  {journal} {\bibinfo
   {journal} {Phys. Rev. E}\ }\textbf {\bibinfo {volume} {103}},\ \bibinfo
  {pages} {022142} (\bibinfo {year} {2021})}\BibitemShut {NoStop}%
\bibitem [{\citenamefont {Feng}\ and\ \citenamefont
  {Skinner}(2021)}]{feng2021_casimir}%
  \BibitemOpen
  \bibfield  {author} {\bibinfo {author} {\bibfnamefont {X.}~\bibnamefont
  {Feng}}\ and\ \bibinfo {author} {\bibfnamefont {B.}~\bibnamefont {Skinner}},\
  }\href@noop {} {\bibinfo {title} {{Hilbert space fragmentation produces a
  "fracton Casimir effect"}}} (\bibinfo {year} {2021}),\ \Eprint
  {https://arxiv.org/abs/2105.11465} {arXiv:2105.11465} \BibitemShut {NoStop}%
\bibitem [{\citenamefont {Chen}\ \emph {et~al.}(2020)\citenamefont {Chen},
  \citenamefont {Gu},\ and\ \citenamefont {A.Lucas}}]{chen2020_slow}%
  \BibitemOpen
  \bibfield  {author} {\bibinfo {author} {\bibfnamefont {X.}~\bibnamefont
  {Chen}}, \bibinfo {author} {\bibfnamefont {Y.}~\bibnamefont {Gu}},\ and\
  \bibinfo {author} {\bibnamefont {A.Lucas}},\ }\bibfield  {title} {\bibinfo
  {title} {{Many-body quantum dynamics slows down at low density}},\ }\href
  {https://doi.org/10.21468/SciPostPhys.9.5.071} {\bibfield  {journal}
  {\bibinfo  {journal} {SciPost Phys.}\ }\textbf {\bibinfo {volume} {9}},\
  \bibinfo {pages} {71} (\bibinfo {year} {2020})}\BibitemShut {NoStop}%
\bibitem [{\citenamefont {Liu}\ \emph {et~al.}(2021)\citenamefont {Liu},
  \citenamefont {Willsher}, \citenamefont {Bilitewski}, \citenamefont {Li},
  \citenamefont {Smith}, \citenamefont {Christensen}, \citenamefont
  {Moessner},\ and\ \citenamefont {Knolle}}]{liu2021_butterfly}%
  \BibitemOpen
  \bibfield  {author} {\bibinfo {author} {\bibfnamefont {S.-W.}\ \bibnamefont
  {Liu}}, \bibinfo {author} {\bibfnamefont {J.}~\bibnamefont {Willsher}},
  \bibinfo {author} {\bibfnamefont {T.}~\bibnamefont {Bilitewski}}, \bibinfo
  {author} {\bibfnamefont {J.-J.}\ \bibnamefont {Li}}, \bibinfo {author}
  {\bibfnamefont {A.}~\bibnamefont {Smith}}, \bibinfo {author} {\bibfnamefont
  {K.}~\bibnamefont {Christensen}}, \bibinfo {author} {\bibfnamefont
  {R.}~\bibnamefont {Moessner}},\ and\ \bibinfo {author} {\bibfnamefont
  {J.}~\bibnamefont {Knolle}},\ }\bibfield  {title} {\bibinfo {title}
  {{Butterfly effect and spatial structure of information spreading in a
  chaotic cellular automaton}},\ }\href
  {https://doi.org/10.1103/PhysRevB.103.094109} {\bibfield  {journal} {\bibinfo
   {journal} {Phys. Rev. B}\ }\textbf {\bibinfo {volume} {103}},\ \bibinfo
  {pages} {094109} (\bibinfo {year} {2021})}\BibitemShut {NoStop}%
\bibitem [{sup()}]{suppmat}%
  \BibitemOpen
  \href@noop {} {\emph {\bibinfo {title} {See {S}upplementary
  {M}aterial}}}\BibitemShut {NoStop}%
\bibitem [{\citenamefont {Hahn}\ \emph {et~al.}(2021)\citenamefont {Hahn},
  \citenamefont {McClarty},\ and\ \citenamefont
  {Luitz}}]{hahn2021_information}%
  \BibitemOpen
  \bibfield  {author} {\bibinfo {author} {\bibfnamefont {D.}~\bibnamefont
  {Hahn}}, \bibinfo {author} {\bibfnamefont {P.~A.}\ \bibnamefont {McClarty}},\
  and\ \bibinfo {author} {\bibfnamefont {D.~J.}\ \bibnamefont {Luitz}},\
  }\href@noop {} {\bibinfo {title} {{Information Dynamics in a Model with
  Hilbert Space Fragmentation}}} (\bibinfo {year} {2021}),\ \Eprint
  {https://arxiv.org/abs/2104.00692} {arXiv:2104.00692} \BibitemShut {NoStop}%
\bibitem [{\citenamefont {Feller}(2008)}]{feller2008_probability}%
  \BibitemOpen
  \bibfield  {author} {\bibinfo {author} {\bibfnamefont {W.}~\bibnamefont
  {Feller}},\ }\href@noop {} {\emph {\bibinfo {title} {{An introduction to
  probability theory and its applications, vol 2}}}}\ (\bibinfo  {publisher}
  {John Wiley \& Sons},\ \bibinfo {year} {2008})\BibitemShut {NoStop}%
\bibitem [{\citenamefont {Shlesinger}(1974)}]{shlesinger1974_random}%
  \BibitemOpen
  \bibfield  {author} {\bibinfo {author} {\bibfnamefont {M.~F.}\ \bibnamefont
  {Shlesinger}},\ }\bibfield  {title} {\bibinfo {title} {{Asymptotic solutions
  of continuous-time random walks}},\ }\href
  {https://doi.org/10.1007/BF01008803} {\bibfield  {journal} {\bibinfo
  {journal} {Journal of Statistical Physics}\ }\textbf {\bibinfo {volume}
  {10}},\ \bibinfo {pages} {421} (\bibinfo {year} {1974})}\BibitemShut
  {NoStop}%
\bibitem [{\citenamefont {Feldmeier}\ \emph {et~al.}(2021)\citenamefont
  {Feldmeier}, \citenamefont {Pollmann},\ and\ \citenamefont
  {Knap}}]{feldmeier2021_fractondimer}%
  \BibitemOpen
  \bibfield  {author} {\bibinfo {author} {\bibfnamefont {J.}~\bibnamefont
  {Feldmeier}}, \bibinfo {author} {\bibfnamefont {F.}~\bibnamefont
  {Pollmann}},\ and\ \bibinfo {author} {\bibfnamefont {M.}~\bibnamefont
  {Knap}},\ }\bibfield  {title} {\bibinfo {title} {{Emergent fracton dynamics
  in a nonplanar dimer model}},\ }\href
  {https://doi.org/10.1103/PhysRevB.103.094303} {\bibfield  {journal} {\bibinfo
   {journal} {Phys. Rev. B}\ }\textbf {\bibinfo {volume} {103}},\ \bibinfo
  {pages} {094303} (\bibinfo {year} {2021})}\BibitemShut {NoStop}%
\end{thebibliography}%

\newpage
\leavevmode \newpage

\setcounter{equation}{0}
\setcounter{page}{1}
\setcounter{figure}{0}
\renewcommand{\thepage}{S\arabic{page}}  
\renewcommand{\thefigure}{S\arabic{figure}}
\renewcommand{\theequation}{S\arabic{equation}}
\onecolumngrid
\begin{center}
\textbf{Supplemental Material:}\\
\textbf{Critically slow operator dynamics in constrained many-body systems}\\ \vspace{10pt}
Johannes Feldmeier$^{1,2}$ and Michael Knap$^{1,2}$ \\ \vspace{6pt}

$^1$\textit{\small{Department of Physics and Institute for Advanced Study, Technical University of Munich, 85748 Garching, Germany}} \\
$^2$\textit{\small{Munich Center for Quantum Science and Technology (MCQST), Schellingstr. 4, D-80799 M{\"u}nchen, Germany}}
\vspace{10pt}
\end{center}
\maketitle
\twocolumngrid

\subsection{A. Shape of out-of-time-order correlations}
\subsubsection{Algebraic tails in the ergodic regime}
\begin{figure}[t]
\centering
\includegraphics[trim={0cm 0cm 0cm 0cm},clip,width=0.95\linewidth]{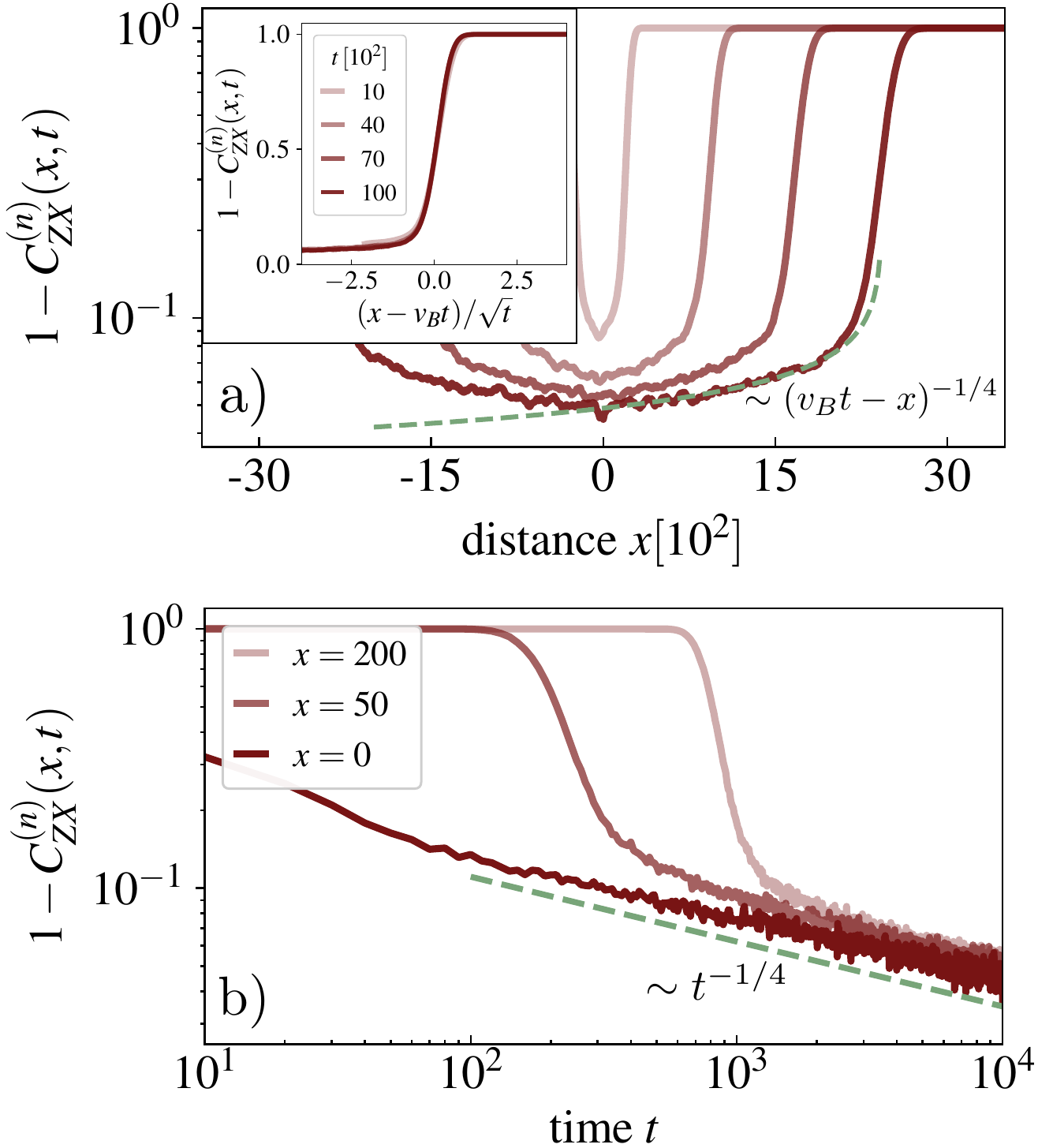}
\caption{\textbf{Algebraic tails from conservation laws.} \textbf{a)} Spatial shape of the $ZX$ -- OTOC at several instances at late times; here for a dipole-conserving circuit with five-state local Hilbert space. The OTOC features an algebraic tail behind the ballistic front. \textbf{b)} In the region where the operator front has passed, the OTOC relaxes algebraically in time $\sim t^{-1/4}$.}
\label{fig:A1}
\end{figure}

As referred to in the main the text, we can analyze the spatio-temporal form of the OTOCs in more detail. In particular, following the considerations of Refs.~\cite{khemani2018_operator,Rakovszky18}, in the ergodic phase the OTOC $C^{(n)}_{ZX}(x,t)$  is expected to feature algebraic tails behind a ballistically propagating front. This is due to the overlap of the $\hat{Z}$ -- operator with the conservation laws in the system, which implies the presence of a conserved operator weight in the Heisenberg time evolution that relaxes only hydrodynamically slow. With the subdiffusive transport properties of Eq.~(1) of the main text, the expected shape of the arising tails is
\begin{equation} \label{eq:A1}
1-C^{(n)}_{ZX}(x,t) \sim \left(v_Bt - x  \right)^{-1/z} = \left(v_Bt - x  \right)^{-1/4} ,
\end{equation}
where $v_B$ is the `Butterfly' velocity, that is the speed of the propagating front. \eq{eq:A1} describes an algebraic relaxation of the OTOC towards its stationary value (which has been normalized to unity here) and should be valid for $v_Bt \gg x$, a long time after the front has passed the point at $x$. While we generally expect the tails \eq{eq:A1} to emerge at very long times on the ergodic side of the localization transition, we note that the presence of waiting times severely affects the dynamics on the numerically accessible timescales even at half-filling $n=1$, as e.g. visible in Fig.~4(a) of the main text. Thus, within our accessible timescales, the chosen three-state local Hilbert space does not lie deep enough in the ergodic phase to verify \eq{eq:A1}. Instead, here, we choose to go even further into the ergodic regime by increasing the local Hilbert space dimension to $5$. \figc{fig:A1}{a} displays the associated spatial shape of the $ZX$ -- OTOC for several instances at late times. We identify a clear tail behind the ballistic front, and the expected relation \eq{eq:A1} provides a good fit to this tail. In addition, we verify in \figc{fig:A1}{b} that the OTOC relaxes in time as $\sim t^{-1/4}$ behind the front as predicted by \eq{eq:A1} as well. Furthermore, we notice that the ballistically moving front itself broadens diffusively according to $1-C^{(n)}_{ZX}(x\approx v_Bt,t) \approx \mathrm{erf}(\frac{x-v_Bt}{\sqrt{t}})$, see the inset of \figc{fig:A1}{a}, where $\mathrm{erf}(\cdot)$ corresponds to the error function. This diffusive broadening is expected to be independent of the system's conserved quantitites and is in agreement with previous results for systems with either no conservation laws or charge conservation only~\cite{nahum2018_operator,Keyserlingk2018,khemani2018_operator,Rakovszky18}.

\subsubsection{Exponential decay in the localized phase}
\begin{figure}[t]
\centering
\includegraphics[trim={0cm 0cm 0cm 0cm},clip,width=0.99\linewidth]{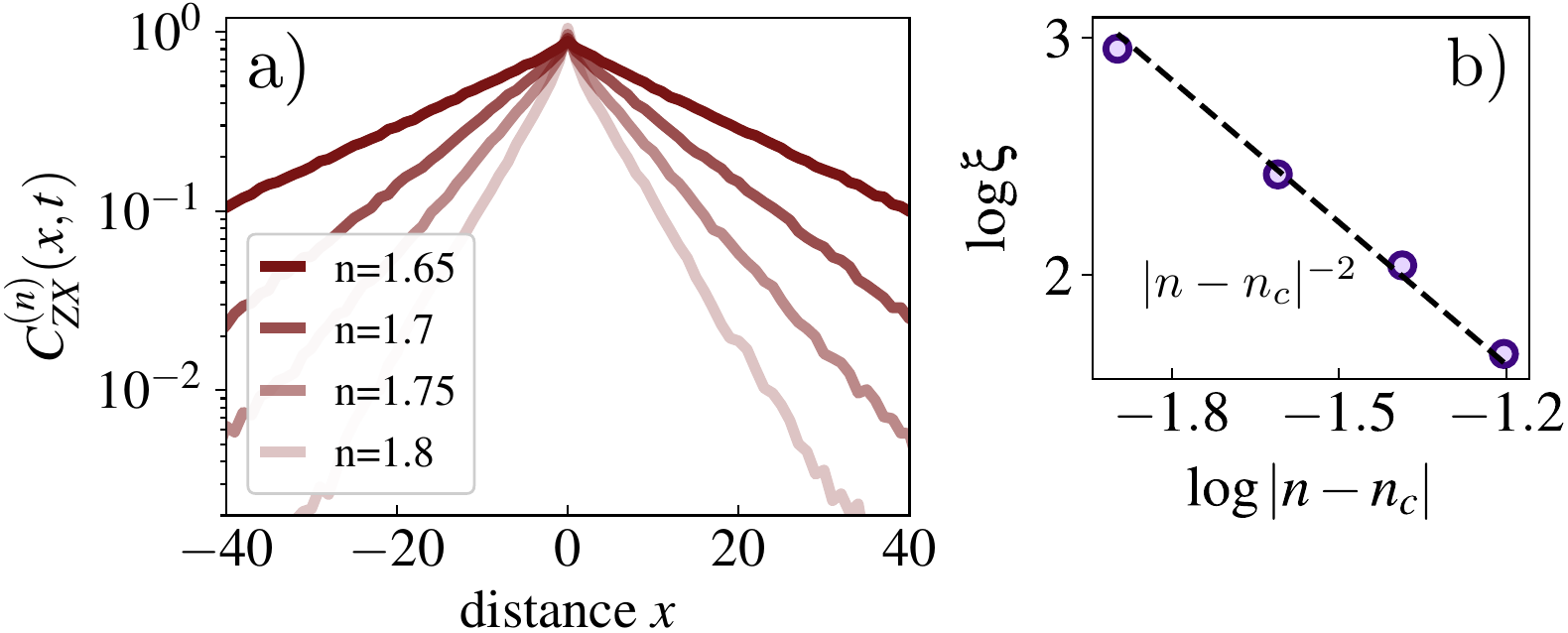}
\caption{\textbf{Exponential shape.} \textbf{a)} In the localized phase $n>n_c$, the OTOC assumes an exponential form. \textbf{b)} The decay is characterized by a correlation length $\xi$ which diverges as $\sim |n-n_c|^{-2}$ towards the critical point.}
\label{fig:A2}
\end{figure}

Similarly to the ergodic regime, we can determine the spatial shape of the OTOC in the localized regime for $n>n_c$. Here, we go back to the three-state local Hilbert space model of the main text and consider densities $n>1.5$. In this regime a finite density of `eternally frozen' sites exists~\cite{morningstar2020_kinetic} whose charge value can never change during the circuit evolution, presenting a hard barrier to the spread of operators. As a consequence, the operator spreading is frozen. The localized phase can then be characterized by a correlation length $\xi \sim |n-n_c|^{-2}$~\cite{morningstar2020_kinetic} that constitutes the relevant length scale in the problem. Hence, we naturally expect the freezing of the OTOC to take place at the scale $\xi$, and further 
\begin{equation} \label{eq:A2}
C^{(n>n_c)}_{ZX}(x,t) \xrightarrow{t\rightarrow \infty} c\, e^{-x/\xi},
\end{equation}
with some constant $c$. We verify \eq{eq:A2} in \figc{fig:A2}{a}, where we show $C^{(n>n_c)}_{ZX}(x,t)$ after convergence at long times and for different densities $n>n_c$. The exponential shape of \eq{eq:A2} is clearly visible and one can also check the expected scaling $\xi \sim |n-n_c|^{-2}$, see \figc{fig:A2}{b}. The exponential form \eq{eq:A2} should be a general feature of localized systems that are characterized by strong Hilbert space fragmentation, as has recently also been observed in~\cite{hahn2021_information}.

\subsection{B. The XX -- OTOC}
While we focused on the $ZX$ -- OTOC above and in the main text, we argue here that the spreading of the front $\overline{\braket{x_{XX}(t)}}$ of the OTOC $C^{(n)}_{XX}(x,t)$ can quite generally be bounded by $\overline{\braket{x_{XX}(t)}} \lesssim 2\overline{\braket{x_{r}(t)}}$, with $\overline{\braket{x_{r}(t)}}$ the front of $C^{(n)}_{ZX}(x,t)$ as considered in the main text.

\begin{figure}[h]
\centering
\includegraphics[trim={0cm 0cm 0cm 0cm},clip,width=0.85\linewidth]{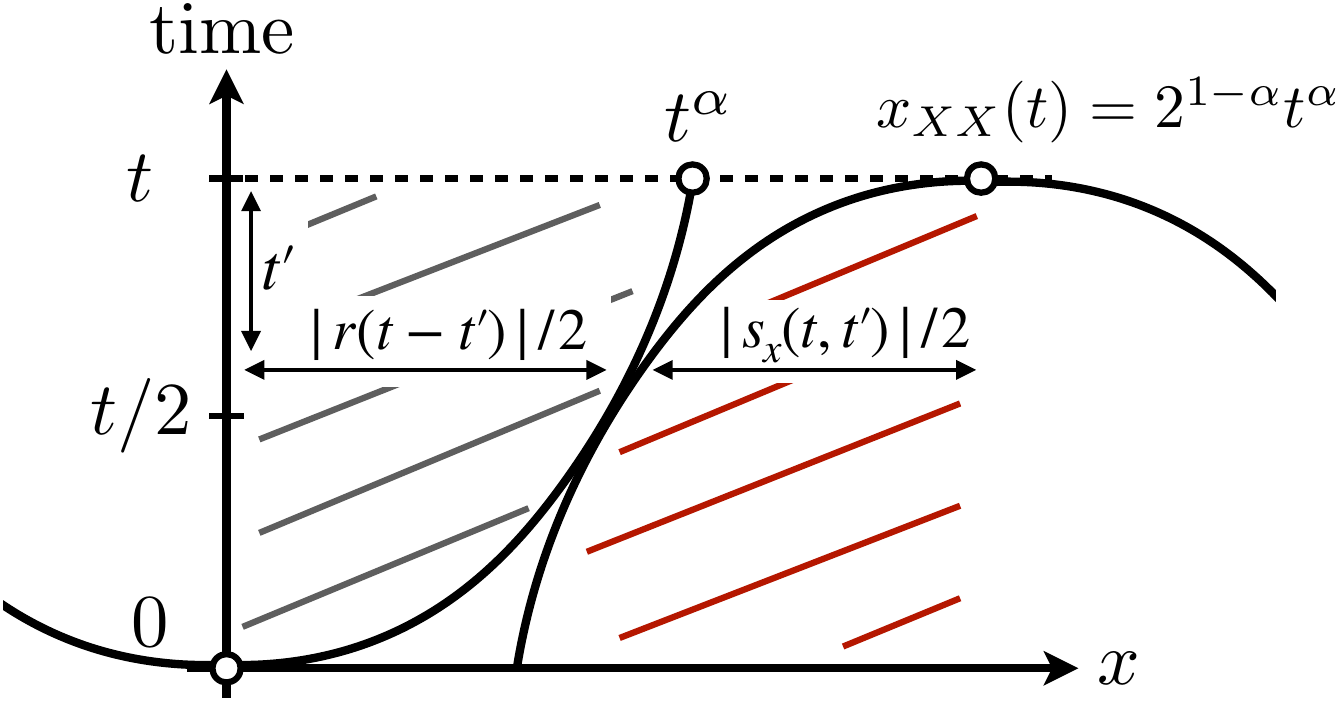}
\caption{\textbf{Bounding the XX -- OTOC.} The forward and backward time evolution, up to $t$ and $t^\prime$ respectively, define two regions $r(t-t^\prime)$ and $s_x(t,t^\prime)$ around the positions $0$ and $x$ where $\hat{X}$-operators have been inserted in \eq{eq:A3}. The front of the OTOC has not yet reached position $x$ when the two regions have no overlap.}
\label{fig:A3}
\end{figure}

To show this, we first write down the expression for the $XX$ --  OTOC,
\begin{equation} \label{eq:A3}
C^{(n)}_{XX}(x,t) = \sum_{\bs{n}} \frac{2e^{-\mu\sum_x n_x}}{Z_{n}} \left( 1 - \overline{\mathrm{Re} \braket{\bs{n}|\hat{X}_x(t)\hat{X}_0 \hat{X}^\dagger_x(t)\hat{X}^\dagger_0|\bs{n}}} \right),
\end{equation}
and consider its out-of-time-ordered part. We notice that for a given $\ket{\bs{n}}$ there is always a region $r(t) \subset \mathbb{Z}$ of size $|r(t)|$ around the origin $x=0$ such that both
\begin{equation} \label{eq:A4}
\begin{split}
\hat{U}(t)\ket{\bs{n}} = \ket{\bs{n}(t)}_{r(t)} \otimes \ket{\bs{n}(t)}_{\bar{r}(t)} \\
\hat{U}(t)\hat{X}_0^\dagger\ket{\bs{n}} = \ket{\bs{n}^\prime(t)}_{r(t)} \otimes \ket{\bs{n}(t)}_{\bar{r}(t)}
\end{split}
\end{equation} 
hold true, where $\bar{r}(t) = \mathbb{Z}\backslash r(t)$ is the complement of $r(t)$ and $\ket{\bs{n}(t)}_{r(t)}$ denotes the restriction of a state $\ket{\bs{n}(t)}$ to the region $r(t)$. According to \eq{eq:A4} the two states only differ inside the region $r(t)$. Furthermore, we are guaranteed the existence of an $x_{XX}(t)>0$ such that for all $x \geq x_{XX}(t)$ and all $t^\prime \in [0,t]$ we can write
\begin{equation} \label{eq:A5}
\begin{split}
\hat{U}^\dagger(t^\prime)\hat{X}_{x}^\dagger\hat{U}(t)\ket{\bs{n}} = &\ket{\bs{n}(t-t^\prime)}_{r(t-t^\prime)} \otimes \ket{\bs{m}(t,t^\prime)}_{s_{x}(t,t^\prime)} \otimes \\
&\otimes \ket{\bs{n}(t-t^\prime)}_{\bar{r}(t-t^\prime)\backslash s_{x}(t,t^\prime)} \\
\hat{U}^\dagger(t^\prime)\hat{X}_{x}^\dagger\hat{U}(t)\hat{X}_0^\dagger\ket{\bs{n}} = &\ket{\bs{n}^\prime(t-t^\prime)}_{r(t-t^\prime)} \otimes \ket{\bs{m}(t,t^\prime)}_{s_{x}(t,t^\prime)} \otimes \\
&\otimes \ket{\bs{n}(t-t^\prime)}_{\bar{r}(t-t^\prime)\backslash s_{x}(t,t^\prime)},
\end{split}
\end{equation}
where $s_{x}(t,t^\prime) \subset \mathbb{Z}$ is some region of size $|s_{x}(t,t^\prime)|$ around $x$, see \fig{fig:A3} for an illustration. It can be verified that for all $x$ for which \eq{eq:A5} holds, $1 - \mathrm{Re} \braket{\bs{n}|\hat{X}^\dagger_{x}(t)\hat{X}^\dagger_0\hat{X}_{x}(t)\hat{X}_0|\bs{n}}  = 0$ in \eq{eq:A3}. Thus, $x_{XX}(t)$ constitutes the boundary of the single-shot OTOC that is associated to $C^{(n)}_{XX}(x,t)$. We emphasize that \eq{eq:A5} will hold for some $t^\prime$ in general only if it also holds for all $t^{\prime\prime} < t^\prime$, i.e. if the two regions $r(t-t^{\prime\prime})$ and $s_{x}(t,t^{\prime\prime})$ do not overlap for any $t^{\prime\prime} < t^\prime$. This is why we are generally required to demand that \eq{eq:A5} hold for all $t^\prime \in [0,t]$ and not only for $t^\prime = t$.

The above requirement that the two regions do not overlap implies that the boundary $x_{XX}(t)$ should fulfill $x_{XX}(t) \approx \frac{1}{2}\left(|r(t-t^\prime)|+|s_{x_{XX}(t)}(t,t^\prime)|\right)$ for all $t^\prime \in [0,t]$ and therefore $x_{XX}(t) \approx \frac{1}{2} \max_{t^\prime}\left[|r(t-t^\prime)|+|s_{x_{XX}(t)}(t,t^\prime)|\right]$. In order to obtain $\overline{\braket{x_{XX}(t)}}$, we now make the following approximations: $1)$ The regions $r(t-t^\prime)$ and $s_{x_{XX}(t)}(t,t^\prime)$ are expected to evolve approximately independently and thus $\overline{\braket{x_{XX}(t)}} \approx \frac{1}{2} \max_{t^\prime}\left[\overline{\braket{|r(t-t^\prime)|}}+\overline{\braket{|s_{x_{XX}(t)}(t,t^\prime)|}}\right]$. $2)$ After averaging, the region $s_{x_{XX}(t)}(t,t^\prime)$ should grow only as a function of $t^\prime$, independently of $t$ and $x_{XX}(t)$ (see \fig{fig:A3}). In particular, it should grow at the same rate as the region $r$. We therefore set $\overline{\braket{|s_{x_{XX}(t)}(t,t^\prime)|}} = \overline{\braket{|r(t^\prime)|}}$.
Using these two approximations as well as the growth $\overline{\braket{|r(t)|}} = 2\overline{\braket{x_r(t)}} =  2 v_Bt^{\alpha}$ of the region $r$ which is given by the boundary of the $ZX$ -- OTOC, we obtain
\begin{equation} \label{eq:A6}
\begin{split}
\overline{\braket{x_{XX}(t)}} &\approx \max_{t^\prime \in [0,t]} \left[\overline{\braket{x_r(t-t^\prime)}} + \overline{\braket{x_r(t^\prime)}} \right] =\\
&= \max_{t^\prime \in [0,t]} \left[ v_B(t-t^\prime)^\alpha + v_B(t^\prime)^\alpha \right] = 2^{1-\alpha}\,v_B t^\alpha =\\
&= 2^{1-\alpha} \overline{\braket{x_r(t)}} \leq 2 \overline{\braket{x_r(t)}}.
\end{split}
\end{equation}
The above argument is expected to hold whenever the $ZX$ -- OTOC is described by an algebraic growth $\overline{\braket{x_r(t)}} =  v_Bt^{\alpha}$ with $\alpha \leq 1$. In particular, we see that the $XX$ --  OTOC spreads with the same algebraic exponent, which concludes our argument.

\end{document}